\documentclass[twocolumn]{aastex631}

\usepackage{newtxtext,newtxmath}
\usepackage{graphicx}	
\usepackage{xcolor}
\usepackage{bm}
\usepackage{anyfontsize}

\shorttitle{variability time scale in jetted AGN}
\shortauthors{Xiong et al.}

\begin{document}

\title{A characteristic optical variability time scale in jetted active galactic nuclei: a large gamma-ray emission sample}

\author[0000-0002-6809-9575]{Dingrong Xiong}
\altaffiliation{Corresponding authors}
\affiliation{Yunnan Observatories, Chinese Academy of Sciences, 396 Yangfangwang, Guandu District, Kunming, 650216, People's Republic of China; \textcolor{blue}{xiongdingrong@ynao.ac.cn}; \textcolor{blue}{baijinming@ynao.ac.cn}}
\affiliation{Center for Astronomical Mega-Science, Chinese Academy of Sciences, 20A Datun Road, Chaoyang District, Beijing, 100012, People's Republic of China}
\affiliation{Key Laboratory for the Structure and Evolution of Celestial Objects, Chinese Academy of Sciences, 396 Yangfangwang, Guandu District, Kunming, 650216, People's Republic of China}

\author[0000-0002-0771-2153]{Mouyuan Sun}
\affiliation{Department of Astronomy, Xiamen University, Xiamen, Fujian 361005, People's Republic of China}

\author[0000-0002-4419-6434]{Jun-Xian Wang}
\affiliation{Department of Astronomy, University of Science and Technology of China, Hefei 230026, People's Republic of China}
\affiliation{School of Astronomy and Space Science, University of Science and Technology of China, Hefei 230026, People's Republic of China}

\author[0000-0002-5929-0968]{Junhui Fan}
\affiliation{Center for Astrophysics, Guangzhou University, Guangzhou 510006, People's Republic of China}
\affiliation{Astronomy Science and Technology Research Laboratory of Department of Education of Guangdong Province, Guangzhou 510006, People's Republic of China}

\author[0000-0002-1935-8104]{Yongquan Xue}
\affiliation{Department of Astronomy, University of Science and Technology of China, Hefei 230026, People's Republic of China}
\affiliation{School of Astronomy and Space Science, University of Science and Technology of China, Hefei 230026, People's Republic of China}

\author[0000-0002-4455-6946]{Minfeng Gu}
\affiliation{Key Laboratory for Research in Galaxies and Cosmology, Shanghai Astronomical Observatory, Chinese Academy of Sciences, 80 Nandan Road, Shanghai 200030, People's Republic of China}

\author[0000-0002-1908-0536]{Liang Chen}
\affiliation{Key Laboratory for Research in Galaxies and Cosmology, Shanghai Astronomical Observatory, Chinese Academy of Sciences, 80 Nandan Road, Shanghai 200030, People's Republic of China}

\author[0000-0001-5895-0189]{Yongyun Chen}
\affiliation{College of Physics and Electronic Engineering, Qujing Normal University, Qujing 655011, People's Republic of China}

\author[0000-0003-1028-8733]{Nan Ding}
\affiliation{School of Physical Science and Technology, Kunming University 650214, People's Republic of China}

\author[0000-0002-7072-2522]{Fei Guo}
\affiliation{Yunnan Normal University, Kunming 650500, People's Republic of China}

\author[0000-0002-7077-7195]{Jirong Mao}
\affiliation{Yunnan Observatories, Chinese Academy of Sciences, 396 Yangfangwang, Guandu District, Kunming, 650216, People's Republic of China; \textcolor{blue}{xiongdingrong@ynao.ac.cn}; \textcolor{blue}{baijinming@ynao.ac.cn}}
\affiliation{Center for Astronomical Mega-Science, Chinese Academy of Sciences, 20A Datun Road, Chaoyang District, Beijing, 100012, People's Republic of China}
\affiliation{Key Laboratory for the Structure and Evolution of Celestial Objects, Chinese Academy of Sciences, 396 Yangfangwang, Guandu District, Kunming, 650216, People's Republic of China}

\author[0000-0002-1497-8371]{Guowei Ren}
\affiliation{Department of Astronomy, University of Science and Technology of China, Hefei 230026, People's Republic of China}
\affiliation{School of Astronomy and Space Science, University of Science and Technology of China, Hefei 230026, People's Republic of China}

\author[0000-0003-1721-151X]{Rui Xue}
\affiliation{Department of Physics, Zhejiang Normal University, Jinhua 321004, People's Republic of China}

\author[0000-0003-4895-1406]{Dahai Yan}
\affiliation{Department of Astronomy, Key Laboratory of Astroparticle Physics of Yunnan Province, Yunnan University, Kunming 650091, People's Republic of China}

\author{Shenbang Yang}
\affiliation{Faculty of Science, Kunming University of Science and Technology, Kunming 650500, People's Republic of China}

\author[0009-0008-1997-1540]{Haiyun Zhang}
\affiliation{Department of Astronomy, Key Laboratory of Astroparticle Physics of Yunnan Province, Yunnan University, Kunming 650091, People's Republic of China}

\author{Jinming Bai}
\altaffiliation{Corresponding authors}
\affiliation{Yunnan Observatories, Chinese Academy of Sciences, 396 Yangfangwang, Guandu District, Kunming, 650216, People's Republic of China; \textcolor{blue}{xiongdingrong@ynao.ac.cn}; \textcolor{blue}{baijinming@ynao.ac.cn}}
\affiliation{Center for Astronomical Mega-Science, Chinese Academy of Sciences, 20A Datun Road, Chaoyang District, Beijing, 100012, People's Republic of China}
\affiliation{Key Laboratory for the Structure and Evolution of Celestial Objects, Chinese Academy of Sciences, 396 Yangfangwang, Guandu District, Kunming, 650216, People's Republic of China}

\begin{abstract}
The variability mechanisms from jetted AGNs are still under debate. Here the damped random walk (DRW) model, implemented through Gaussian Processe (GPs), is used to fit the $ZTF$ long-term optical light curves of 1684 $\gamma$-ray emission jetted AGNs. This analysis yields one of the largest samples with characteristic optical variability timescales for jetted AGNs. A single DRW model from GPs can fit the optical light curve of most jetted AGNs well/potentially well, while there are still some jetted AGNs whose light curve can not be fitted well by a single DRW model. After the jet power, proxied by gamma-ray luminosity, is introduced as a new parameter, new relationships among intrinsic variability time scales, black hole mass and jet power are discovered for efficient accretion AGNs ($\tau^{\rm in} \propto M_{\rm BH}^{0.29^{+0.06}_{-0.06}}P_{\rm jet}^{-0.3^{+0.03}_{-0.03}}$ with scatter of approximately 0.09~dex) and for inefficient accretion AGNs ($\tau^{\rm in} \propto M_{\rm BH}^{0.06^{+0.07}_{-0.07}}P_{\rm jet}^{0.37^{+0.11}_{-0.11}}$ with scatter of approximately 0.14~dex), respectively. Our results support that the optical variability of jetted AGNs with efficient accretion may originate within the standard accretion disk at UV emitting radii similar to non-jetted AGNs, and is directly related to the acceleration of shock in the jet and then enhanced through the beaming effect in beamed AGNs. For the jetted AGNs with inefficient accretion, the intrinsic timescale is consistent with the escape timescale of electrons.

\end{abstract}


\section{Introduction} \label{sec:intro}

Variability is a defining property of active galactic nuclei (AGNs), making it a fundamental tool for understanding the AGN engine and mapping its immediate environment \citep{Padovani2017}. 
Despite its importance, the physical driver of variability across different AGN types remains an outstanding question, actively debated in the literature \citep[e.g.,][]{McHardy2006, Kording2007, Cackett2007, Valtonen2008, Bottcher2010, MacLeod2010, Risaliti2011, Mushotzky2011, Zuo2012, Marscher2014, Kelly2014, Finke2014, SunYH2014, Cai2016, Xiong2017, Cai2018, Sun2020,Bhatta2021, SouH2022,Ding2024, WuL2024}. This long-standing problem motivates continuous exploration into variability mechanisms across AGN subclasses.

Jetted AGNs, traditionally known as radio-loud AGNs, represent a unique subset of AGNs, whose variability offers valuable insights into both the central accretion disk and the relativistic jet, making them a focus of particular interest. These systems include beamed blazars (flat-spectrum radio quasars and BL Lacertae object, i.e., FSRQ and BL Lac), as well as unbeamed radio galaxies (Fanaroff–Riley type I and II radio galaxies, i.e., FRI and FRII), steep-spectrum radio quasars (SSRQ), and jetted narrow-line Seyfert 1 galaxy (NLSY1) \citep[e.g.,][]{Urry1995,Abdo2009,Falomo2014,Padovani2017,Foschini2011}. 

Among these, blazars, a rare subclass with jet closely aligned with the observer's line of sight, exhibit strong relativistic beaming effects which significantly shorten the observed variability timescales and amplify the luminosity by several orders of magnitude \citep[e.g.,][]{Urry1995,Padovani2017}. In these sources, synchrotron radiation from the jet dominates the emission across radio to optical wavelengths, with accretion disk contributions often being negligible, particularly in radiatively inefficient systems \citep[e.g.,][]{Ghisellini2009,Giommi2012,Williamson2014,Falomo2014}. 
Radio galaxies, with jets oriented at larger angles relative to the observer's line of sight, are considered the non-beamed counterparts of blazars. According to the unified model of AGNs suggesting that FRII radio galaxies are the parent population of FSRQs and FRI radio galaxies are the parent population of BL Lacs. 
Consequently, after correcting for the beaming effect in blazars, radio galaxies and blazars are expected to exhibit similar physical properties, such as variability.
Additionally, SSRQs are suggested to be the parent population of jetted NLSY1 \citep[e.g.,][]{Berton2016}. For jetted AGNs, a physical origin classification has also been proposed, distinguishing sources by different accretion rates
\citep[e.g.,][]{Ghisellini2001,Ghisellini2011,Sbarrato2012,Xiong2014,Padovani2017}.

Jetted AGNs exhibit variability across a wide range of wavelengths, from the radio to TeV energies, and over diverse timescales, from minutes to years \citep[e.g.,][]{Wagner1995,Urry1995,Bai1998,Fan2005,Falomo2014,Padovani2017}. Various mechanisms have been proposed to explain the  variability in jetted AGNs, including particle injection, acceleration, and cooling within the jet, with possible intervention of shock waves (or turbulence) \citep[e.g.,][]{Marscher1985,Sikora2001,Ghisellini2002,Marscher2014}, geometric interpretation or changes in jet regions \citep[e.g.,][]{Villata1999,Marscher2008,Abdo2010,Raiteri2017,Xiong2020,2022PhRvD.105b3005W, 2023MNRAS.526.5054L,Raiteri2024}, accretion disk-related processes \citep[e.g.,][]{Gu2006,Zhang2024}, gravitational microlensing \citep[e.g.,][]{Torres2003}, kink instabilities, and magnetic reconnection \citep[e.g.,][]{Jorstad2022,Ding2019,Chang2024}.

The power spectral density (PSD) has been widely used for variability studies. A break in the PSD may indicate a characteristic variability timescale; however, the estimated value can be affected by distortions caused by aliasing and red noise leakage \citep[e.g.,][]{Uttley2002,Burke2021}. 
In contrast, Gaussian Processes (GPs) have become an increasingly important and powerful tool for modeling stochastic signals in time-domain astronomy
\citep{Aigrain2023}. The damped random walk (DRW) model is the simplest member of a family of continuous auto-regressive with moving average (CARMA) models for GPs \citep[e.g.,][]{Burke2021,Yang2021}. 

Fitting the optical light curves with DRW has become a common method for describing the variability of AGNs \citep[e.g.,][]{Kelly2009,MacLeod2010,Ruan2012,Zu2013,Suberlak2021,Burke2021,Zhang2022,Zhou2024,Ren2024, Zhang2024}. Analyzing the damping timescale and its correlation with physical parameters, such as black hole mass, wavelength, and accretion rate, can provide valuable insights into the underlying variability mechanisms \citep[e.g.,][]{Kelly2009,MacLeod2010,Suberlak2021,Burke2021,Zhang2022,Zhou2024,Ren2024, Zhang2024}.
For instance, \citet{Burke2021} used GPs regression to fit a DRW model to optical light curves of non-jetted AGNs. They found a correlation between the damping time scale and black hole mass, with the damping time scale aligning with the expected thermal timescale at the ultraviolet-emitting radius, as predicted by standard accretion disk theory.

Building on this, researchers have applied the DRW model using GPs regression to $\gamma$-ray, optical, X-ray light curves of jetted AGNs \citep[e.g.,][]{Yang2021, Zhang2022, Zhang2023, Zhang2024}. 
However, a widely accepted physical interpretation of the optical variability timescale in jetted AGNs remains elusive, due to factors such as sample size limitations, beaming effects, short observation periods, and methodological constraints.

In this study, we apply GPs to fit DRW model to long-term optical light curves for a large gamma-ray-emitting AGN sample, with the aim of exploring the physical interpretation of optical variability timescales in jetted AGNs.
After considering jetted AGNs with well-fitted DRW models and correcting for the beaming effect, we uncover new relationships between the damping time scale and other physical parameters, specifically for efficient and inefficient accretion AGNs. 

The structure of this paper is as follows: Section 2 describes the sample selection and methodology. The main results are presented in Section 3. Section 4 discusses the implications of our findings, while Section 5 summarizes the conclusions. Throughout this work, we adopt a flat $\Lambda$CDM cosmology with $H_{0}=67.7~{\rm km~s^{-1}~Mpc^{-1}}$, $\Omega_{\rm m}=0.31$, and $\Omega_{\rm \Lambda}=0.69$ \citep{Planck2020}.

\begin{deluxetable}{cccccccccccccccc}
\tablenum{1}
\tablecaption{The fitting results of the damped random walk (DRW) model using Gaussian Processes (GPs) for $\gamma$-ray-emitting jetted AGNs. Column (1): source name of 4FGL ; Column (2-4): logarithm of variability damping time scale and its corresponding lower and upper errors; Column (5-7): logarithm of variability amplitude and its corresponding lower and upper errors; Column (8): mean cadence; Column (9): baseline; Column (10): the level of noise $\sigma_{\rm s/n}^2=\sigma_{\rm n}^2+<{\rm err}>^2$; Column (11): the $P$ value of KS test; Column (12): the 95\% confidence limit of the white noise; Column (13): the maximum ACF value among all time lags of residuals; Column (14): the minimum $P$ value of LB test; Column (15): class of jetted AGNs in which bll is BL Lac, fsrq is FSRQ, bcu is blazar of uncertain type, nlsy1 is NLSY1, sey is Seyfert galaxy, ssrq is SSRQ, rdg is radio galaxy, css is compact steep spectrum radio source and agn is AGN; Column (16): name of associated counterpart in other bands. The second row in the header is the unit. This table is available in its entirety in machine-readable form.\label{tab:1}} 
\tablehead{
\colhead{4FGL name} & \colhead{${\rm Log}~\tau$} &  \colhead{$\tau_{\rm le}$} & \colhead{$\tau_{\rm ue}$}  &  \colhead{${\rm Log}~\sigma$} &\colhead{$\sigma_{\rm le}$} & \colhead{$\sigma_{\rm ue}$}  &  \colhead{Cad.} &  \colhead{Bas.}&  \colhead{$\sigma_{\rm s/n}^2$}&  \colhead{$P_{\rm nor}$}&\colhead{Cl.} &\colhead{${\rm ACF}_{\rm max}$} &\colhead{$P_{\rm LB}^{\rm min}$} & \colhead{Class} &\colhead{Name of associated counterpart}\\
\colhead{} & \colhead{(days)} &  \colhead{(days)} & \colhead{(days)}  &  \colhead{(mag)} &\colhead{(mag)} & \colhead{(mag)}  &  \colhead{(days)} &  \colhead{(days)}&  \colhead{(mag)}&  \colhead{}&\colhead{} &\colhead{}&\colhead{} & \colhead{} &\colhead{}
}
\rotate
\startdata
J0001.2-0747		&	3.043	&	-0.381	&	0.622	&	-0.237	&	-0.092	&	0.153	&	0.858	&	1839	&	0.0015	&	0.265	&	0.122	&	0.122	&	0.401	&	bll	&	PMN J0001-0746	\\
J0001.4-0010		&	1.429	&	-0.244	&	0.327	&	-0.514	&	-0.032	&	0.035	&	0.947	&	1843	&	0.0102	&	0.505	&	0.135	&	0.131	&	0.222	&	bll	&	FBQS J0001-0011	\\
J0001.8-2153		&	1.466	&	-0.149	&	0.186	&	-0.134	&	-0.03	&	0.039	&	1.074	&	1602	&	0.0086	&	0.799	&	0.168	&	0.206	&	0.048	&	bcu	&	PKS 2359-221	\\
J0002.3-0815		&	2.077	&	-0.192	&	0.279	&	-0.254	&	-0.041	&	0.063	&	0.893	&	1839	&	0.0077	&	0.675	&	0.127	&	0.152	&	0.261	&	bcu	&	WISEA J000236.06-081532.4	\\
J0003.2+2207		&	1.027	&	-0.246	&	0.273	&	-0.665	&	-0.038	&	0.034	&	0.676	&	1857	&	0.0042	&	0.4	&	0.098	&	0.095	&	0.419	&	bll	&	2MASX J00032450+2204559	\\
J0003.3-1928		&	1.91	&	-0.175	&	0.249	&	-0.109	&	-0.039	&	0.056	&	1.002	&	1631	&	0.0122	&	0.884	&	0.153	&	0.141	&	0.051	&	bcu	&	PKS 0000-197	\\
J0003.5+0717		&	0.518	&	-0.153	&	0.729	&	-0.474	&	-0.049	&	0.017	&	0.817	&	1852	&	0.0045	&	0.053	&	0.116	&	0.101	&	0.81	&	bcu	&	GB6 J0003+0717	\\
J0004.0+0840		&	1.899	&	-0.207	&	0.269	&	-0.275	&	-0.038	&	0.055	&	0.842	&	1851	&	0.0169	&	0.645	&	0.119	&	0.113	&	0.56	&	bll	&	SDSS J000359.23+084138.1	\\
J0004.3+4614		&	1.42	&	-0.104	&	0.123	&	-0.169	&	-0.022	&	0.027	&	0.543	&	1710	&	0.0157	&	0.351	&	0.088	&	0.1	&	0.089	&	fsrq	&	MG4 J000421+4615	\\
\enddata
\end{deluxetable}   
\section{Sample and method} \label{sec:sam}

Launched on 11 June, 2008, the $Fermi$ Large Area Telescope ($LAT$) is an high-energy gamma-ray observatory that covers an energy range from below 20 MeV to over 300 GeV \citep{Atwood2009}. For our study, we utilized the fourth catalog of AGNs detected by $Fermi~LAT$ \citep[4LAC-DR3;][]{Ajello2022}, focusing on the high-latitude sample ($\lvert b \lvert >10^{\circ}$). This catalog is derived from the third data release of the 4FGL catalog \citep[4FGL-DR3;][]{Abdollahi2022} based on 12 years of E $>$ 50 MeV $\gamma$-ray data.  The 4LAC-DR3 high-latitude sample includes 3407 $\gamma$-ray emission AGNs: 755 FSRQs, 1379 BL Lac objects, 1208 blazars of unknown type (BCU), and 65 non-blazar AGNs. The Fermi collaborations employed Bayesian and the likelihood-ratio methods to associate 4LAC sources with counterparts detected at other wavelengths \citep{Ajello2020}. The coordinates of these counterparts were provided in Table A1 from 4LAC-DR3.

In this work we collected optical light curves for these Fermi detected jetted AGNs from the Zwicky transient facility ($ZTF$), which surveyed the northern sky in $g$, $r$ and $i$ bands (with an average cadence of three days) using a 47 deg$^2$ wide-field imager mounted on a 48-inch Schmidt telescope at Mount Palomar \citep{Bellm2019}. The long-term optical light curves were retrieved from the 19th $ZTF$ public data release \footnote{https://irsa.ipac.caltech.edu/data/ZTF/docs/releases/dr19/ztf\_release\_not\\es\_dr19.pdf} \citep{Masci2019}, covering observations from March 2018 through July 2023.

 Making use of ztfquery tool \citep{Rigault2018}, we retrieved $ZTF$ light curves by searching within a 1.5\arcsec\ radius around the coordinates of counterparts. The fractional variability was estimated as described by \citet{Romero1999}, and the distributions are presented in Fig. 1. The average variability amplitudes in the $g$-band are slightly greater than those in the $r$-band, with values of $\langle \text{A}_g \rangle = 7.24\%$ and $\langle \text{A}_r \rangle = 6.95\%$. The variability behavior between the two bands does not exhibit significant differences. The light curves in the $g$-band were utilized in the subsequent analysis.

\begin{figure}
    \centering
    \includegraphics[width=0.52\textwidth]{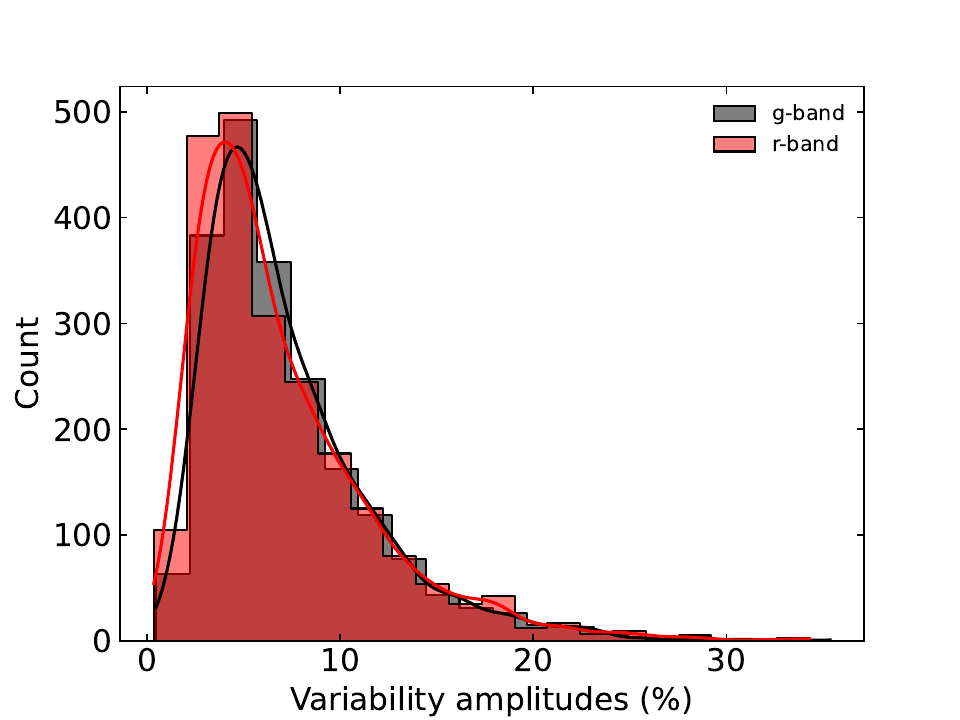}
    \caption{The histograms of variability amplitudes. The lines of different colors represent the kernel density estimates.}
\end{figure}

Reliable light curves were selected based on the following criteria:

1. Each source had more than 25 observations ($N>25$).

2. Outliers with the brightness value exceeded the average by more than 3$\sigma$ were removed. To minimize the unnecessary removal of flare data points, an additional criterion was applied: if more than five data points were removed and they did not conform to white noise, these outliers will be retained. Poor quality data points with too large uncertainties were excluded. The uncertainty threshold was set at 0.22 magnitude, determined as the mean errors on the entire sample plus 3 times the standard deviation of the errors ($<$err$>$+3$\times\sigma_{\rm err}$).

3. Data points with a catflags score = 0 were selected to ensure reliable observations from $ZTF$.  

4. The maximum sampling gap for each light curve was required to be less than one-third of the light curve's baseline (length). This restriction minimized the impact of gaps on the analysis results.

5. The light curves were not dominated by white noise, e.g., the statistic p-value of Ljung-Box (LB) test\footnote{https://www.statsmodels.org/dev/generated/statsmodels.stats.diagnostic.\\acorr\_ljungbox.html} less than 0.05.

6. Following the approach of \citet{Negi2022}, we used only the light curves corresponding to the observation ID with the maximum number of data points. We avoided combining light curves from different observation IDs for the same source to prevent introducing spurious variability \citep{van2021}.

Using these stringent criteria, a total of 1684 $\gamma$-ray emitting AGNs were selected. The results of the DRW model fitting, performed using Gaussian Processes (GPs),  are presented in Table 1.

\begin{figure*}
\centering
\gridline{\fig{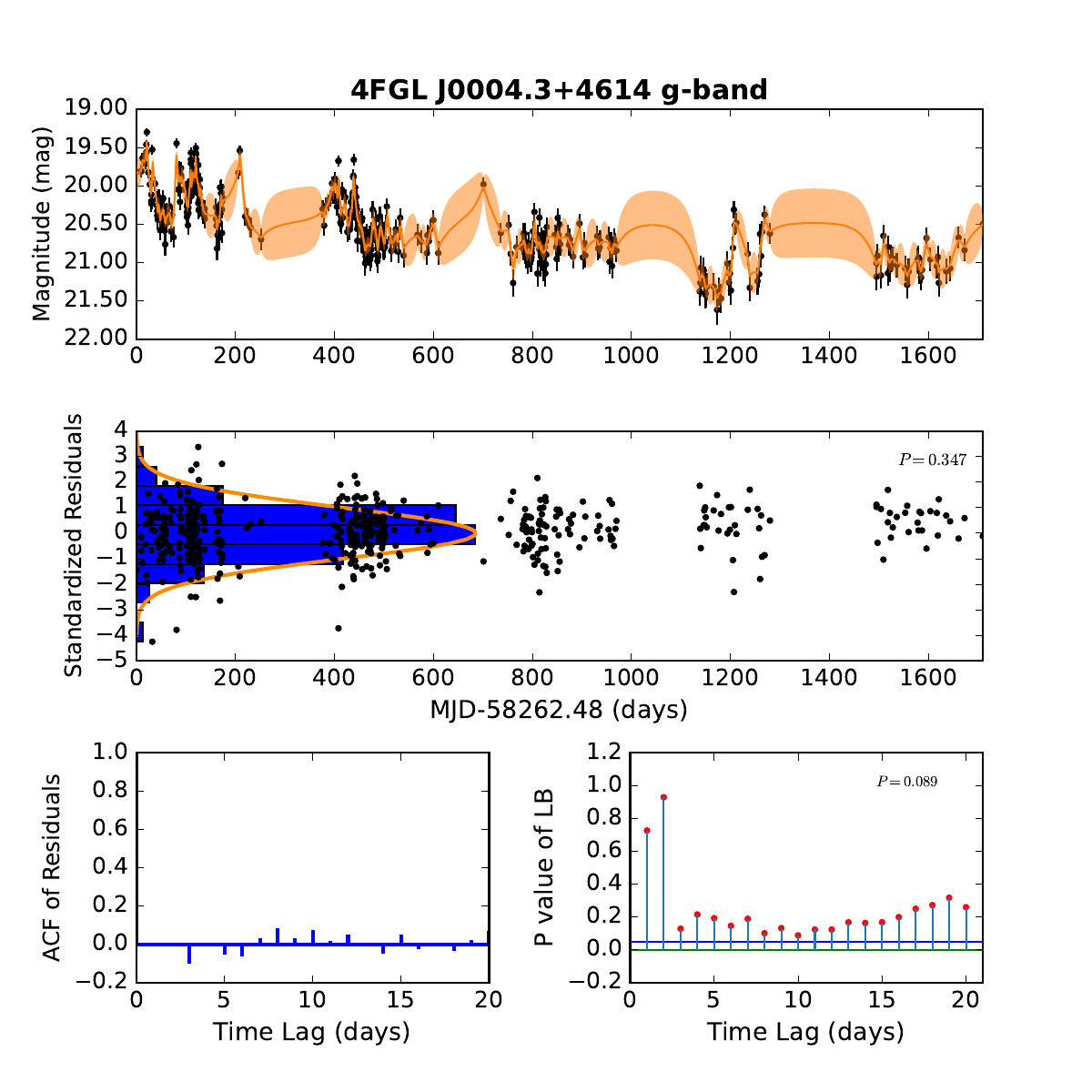}{0.5\textwidth}{(a) Gaussian Processes (GPs) fitting of the damped random walk (DRW) model to the optical light curves and evaluation of fitting results.} \label{fig:1e}
          \fig{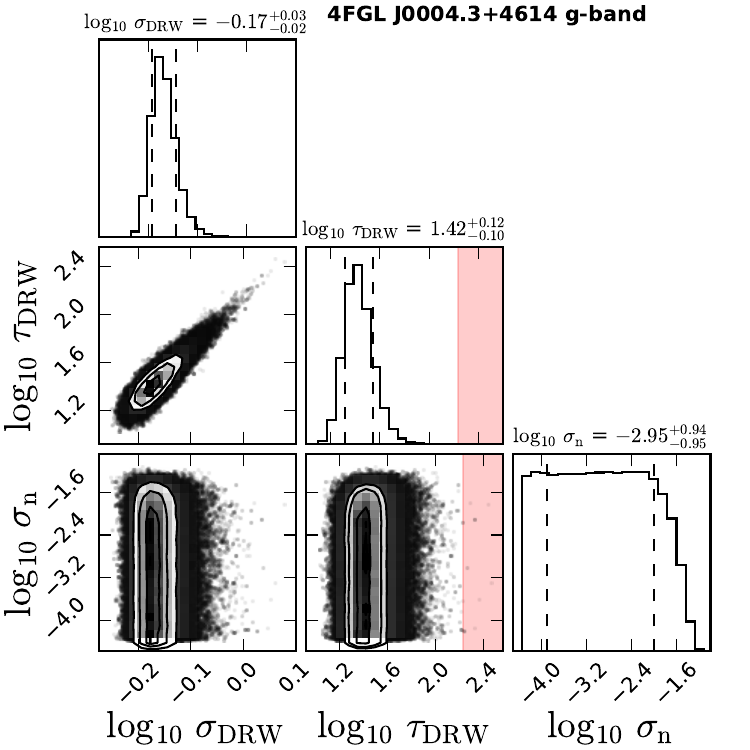}{0.5\textwidth}{(b) The posterior distribution of mode parameters and their covariance.}\label{fig:1f}}
    \caption{Fittings, results, and posterior distribution of parameters. In the top panel of Fig. 2a, the black data points represent the optical light curves, while the orange line and shaded region indicate the model fitting line and the 1$\sigma$ confidence interval, respectively. In the middle panel of Fig. 2a, the standardized residuals (black points) are presented alongside the probability density of standardized residuals (blue histogram) and the best-fit normal distribution (yellow solid line). The $P$ value of the KS test, located in the upper right corner and greater than 0.05, suggests that the standardized residuals follow a normal distribution. The ACF of the residuals (shown with blue lines) are depicted in the bottom left column of Fig. 2a. In the bottom right column, the highest red dot on the vertical axis represents the $P$ value from the LB test corresponding to different time lags, while the horizontal blue line indicates the threshold value of 0.05. The $P$ value in the upper right corner of this column indicates the smallest $P$ value among all time lags considered. In the Fig. 2b, the vertical dashed lines represent the posterior distribution's 16th and 84th percentiles, with the red shaded regions corresponding to timescales greater than 10\% of the light curve length. The PSD of the DRW, as computed by $celerite$, is shown in Fig. A1 of the Appendix.}
    \label{fig:spectra_compare}
\end{figure*} 

\subsection{GPs regression method} \label{subsec:GP}

The $celerite$ algorithm is designed for fast and scalable GPs regression in one dimension, particularly for efficiently evaluating the marginalized likelihood of a dataset under a GP model \citep{Foreman2017}. We utilized the $celerite$ package\footnote{https://celerite.readthedocs.io/en/stable/} to perform GPs regression, fitting light curves with the DRW model, which is the simplest term available in the $celerite$. Previous researches have indicated that the $celerite$ package is effectively applicable to the study of AGN variability \citep[e.g.,][]{Burke2021,Yang2021,Zhang2021,Zhang2022,Zhang2023,Zhang2024,Zhou2024,Ren2024}. 

The kernel function that includes both the DRW model as a real term and a white noise term in the $celerite$, is expressed as follows:
\begin{equation}
k(t_{\rm nm})=ae^{-ct_{\rm nm}}+\sigma^2_{\rm n}\delta_{\rm nm}
=2\sigma^2_{\rm DRW}e^{-t_{\rm nm}/\tau_{\rm DRW}}+\sigma^2_{\rm n}\delta_{\rm nm},
\end{equation}
where $\sigma_{\rm n}$ represents the excess white noise amplitude, $\delta_{\rm nm}$ denotes the Kronecker delta function, and $t_{\rm nm}$ is the time lag between measurements $m$ and $n$. The damping timescale is given by $\tau_{\rm DRW}=1/c$, and the variability amplitude by $\sigma_{\rm DRW}=\sqrt{(a/2)}$. 

The Power Spectral Density (PSD) is formulated as  \citep{Foreman2017}
\begin{equation}
S(\omega)=\sqrt{\frac{2}{\pi}}\frac{a}{c}\frac{1}{1+(\omega/c)^2}.
\end{equation}
The PSD of the DRW displays a broken power-law form, with the index transitioning from 0 at low frequencies to -2 at high frequencies, and the broken frequency written as \citep{Zhang2022}
\begin{equation}
f_b=\frac{1}{2 \pi \tau_{\rm DRW}}.
\end{equation}
 The findings from \citet{Zhang2022} supported that the variability PSDs of $\gamma$-ray-emitting AGNs conform to the standard DRW PSD form.

The $celerite$ package allowed us to execute the following steps for the fitting process:

1. Established the GP model (real term + white noise term);

2. Defined a cost function and employed the L-BFGS-B non-linear optimization routine to identify the maximum likelihood parameters;

3. Selectd priors and combined them with our likelihood function to compute the log probability;

4. Fitted the model using emcee \citep{Foreman2013}, which included initializing the walkers and conducting both burn-in and production chains;

5. Obtained chains of posterior samples and plotted the posterior distribution of model parameters.

To assess model fit quality, we analyzed the standardized residuals with respect to Gaussian models \citep{Yang2021}. A good model fit is indicated if the standardized residuals conform to a normal distribution with a mean of zero and a standard deviation of one (i.e., $p$ value of the Kolmogorov-Smirnov (KS) test greater than 0.05) and exhibit characteristics of a Gaussian white-noise series. To determine whether the standardized residuals satisfy the properties of a white noise series, both the auto-correlation function (ACF) and the Ljung-Box (LB) tests were used. A total of 20 time lags were considered for the ACF and LB tests, complying with the common criterion $\text{min}(20, N-1)$, where $N$ is the length of the time series \citep{Box1994}. Meeting both of the following conditions simultaneously indicates that the residuals conform to a white noise series: (1) ACF absolute values for all time lags fall within the 95\% white noise confidence interval (absolute values $CI$), and (2) the LB test yields $p>0.05$ for all time lags. 

There are two additional considerations: ACF absolute values for a few time lags may slightly exceed the 95\% white noise confidence interval $CI$, and the $p$ values from the LB test for a few time lags may be marginally below 0.05 (see Fig. A2a, A2b in the Appendix). When both of the following additional conditions were met, the residuals were also considered to conform to a white noise series: (1) the absolute values of the ACF of residuals for all time lags were less than $ CI + 0.03 $, and (2) the $p$ value from the LB test exceeded 0.02 ($p - 0.03 $). The criterion value of 0.03 was adopted to ensure it was neither too far from nor too close to 0.05; a value too distant from 0.05 could create less stringent conditions, while a value too near 0.05 would provide results nearly indistinguishable from those at 0.05.

The $p$ values obtained from the normal distribution test were slightly below 0.05 (specifically, $p=0.02$), suggesting that, while the residuals did not perfectly conform to a normal distribution, they did not deviate significantly from it. In cases where the model fitting was potentially good, the $p$ values of the normal distribution test exceeded 0.02 and satisfied at least one of the following conditions: (1) the absolute values of the ACF of the standardized residuals were less than $CI + 0.03 $, and (2) the $p$ values from the LB test for all time lags exceeded 0.02. If the standardized residuals failed to meet the criteria for white noise (specifically, if any ACF absolute values exceeded $CI + 0.03$ and any $p$ value from the LB test dropped below 0.02), or if the distribution was not normal (i.e., $p < 0.02$), the model fitting was deemed to be poor (see Fig. A2c and A2d in the Appendix).

Based on these criteria, 922 $\gamma$-ray emission AGNs exhibited good model fits, 345 displayed potentially good model fits, and 417 were categorized as having poor model fits. These results suggest that a single DRW, utilizing GPs regression, does not adequately account for the light curves of all jetted AGNs.

In the subsequent sections, we focus primarily on the jetted AGNs with good model fits. To obtain reliable measurements of the damping timescale, we established three criteria, adapted from \citet{Burke2021}:

1. the damping timescale $\tau_{\rm DRW}< 0.1\times$baseline,

2. the damping timescale $\tau_{\rm DRW} >$ mean cadence,

3. the variability amplitude $\sigma_{\rm DRW}^2 > \sigma^2_{\rm n}+<{\rm err}>^2$.

Ultimately, 606 of the 922 jetted AGNs with good model fitting met these three criteria. An example (the blazar in the first row of Table 2) is illustrated in Fig. 2, with the corresponding PSD of the DRW displayed in Fig. A1 of the Appendix.

\begin{deluxetable*}{ccccccccccl}
\tablenum{2}
\tablewidth{0pt}
\tablecaption{The sample of jetted AGNs with efficient accretion that exhibit good model fitting. Column (1): source name of 4FGL; Column (2): the redshift; Column (3): class of jetted AGNs; Column (4-6): logarithm of observational variability damping time scale and its corresponding lower and upper errors; Column (7): logarithm of black hole mass and its corresponding uncertainty; Column (8): logarithm of accretion disk luminosity and its corresponding uncertainty; Column (9): logarithm of observed $\gamma$-ray luminosity and its corresponding uncertainty; Column (10): $\gamma$-ray photon index when fitting with power law and its corresponding uncertainty; Column (11): Doppler factor and its corresponding uncertainty. The second row in the header is the unit. This table is available in its entirety in machine-readable form.\label{tab:2}} 
\tablehead{
\colhead{4FGL name} & $z$ &\colhead{Class} &\colhead{${\rm Log}~\tau$} &  \colhead{$\tau_{\rm le}$} & \colhead{$\tau_{\rm ue}$}  &  \colhead{${\rm Log}~M_{\rm BH}$} &\colhead{${\rm Log}~L_{\rm disk}$} & \colhead{${\rm Log}~L_{\rm \gamma}^{\rm obs}$}  &  \colhead{${\rm PL}_{\rm Index}$} &  \colhead{$\delta$}\\
\colhead{} & \colhead{} &  \colhead{} & \colhead{(days)}  &  \colhead{(days)} &\colhead{(days)} & \colhead{($M_{\odot}$)}  &  \colhead{(${\rm erg~s^{-1}}$)} &  \colhead{(${\rm erg~s^{-1}}$)}&  \colhead{}&  \colhead{}
}
\startdata
J0004.3+4614	&	1.81	&	fsrq	&	1.42	&	-0.104	&	0.123	&	8.36	$\pm$	0.1	&	46.07	$\pm$	0.03	&	47.35	$\pm$	0.06	&	2.59	$\pm$	0.07	&	7.75	$\pm$	0.59	 \\
J0011.4+0057	&	1.49	&	fsrq	&	1.337	&	-0.102	&	0.12	&	8.66	$\pm$	0.05	&	45.71	$\pm$	0.02	&	47.18	$\pm$	0.04	&	2.35	$\pm$	0.05	&	11.03	$\pm$	0.59	 \\
J0013.6-0424	&	1.07	&	fsrq	&	1.524	&	-0.151	&	0.18	&	7.82	$\pm$	0.08	&	45.03	$\pm$	0.08	&	46.2	$\pm$	0.1	&	2.13	$\pm$	0.17	&	27.16	$\pm$	0.59	 \\
J0030.6-0212	&	1.8	&	fsrq	&	1.078	&	-0.091	&	0.102	&	8.66	$\pm$	0.28	&	45.98	$\pm$	0.02	&	47.81	$\pm$	0.02	&	2.39	$\pm$	0.03	&	7.44	$\pm$	1.22	 \\
J0036.9+1832	&	1.59	&	bcu	&	1.963	&	-0.241	&	0.317	&	8	$\pm$	0.09	&	45.7	$\pm$	0.04	&	47	$\pm$	0.06	&	2.43	$\pm$	0.09	&	5.89	$\pm$	1.22	 \\
J0038.2-2459	&	0.49	&	fsrq	&	1.744	&	-0.16	&	0.219	&	8.14	$\pm$	0.23	&	44.97	$\pm$	0.07	&	46.44	$\pm$	0.01	&	2.24	$\pm$	0.02	&	5.64	$\pm$	1.22	 \\
J0102.4+4214	&	0.87	&	fsrq	&	1.043	&	-0.181	&	0.244	&	8.33	$\pm$	0.24	&	45.66	$\pm$	0.16	&	46.56	$\pm$	0.04	&	2.67	$\pm$	0.06	&	24.26	$\pm$	0.59	 \\
J0152.2+2206	&	1.32	&	fsrq	&	1.723	&	-0.133	&	0.172	&	8.45	$\pm$	0.06	&	46.17	$\pm$	0.02	&	47.03	$\pm$	0.04	&	2.56	$\pm$	0.06	&	4.32	$\pm$	0.59	 \\
J0222.0-1616	&	0.69	&	fsrq	&	1.169	&	-0.1	&	0.117	&	7.58	$\pm$	0.48	&	45.34	$\pm$	0.05	&	46.35	$\pm$	0.03	&	2.38	$\pm$	0.06	&	14.95	$\pm$	0.59	 \\
\enddata
\end{deluxetable*}

\subsection{Correction of beaming effect} \label{subsec:beam}

The $\gamma$-ray emitting blazars are likely to exhibit stronger beaming effects and possess more powerful jets compared to non-$\gamma$-ray emitting blazars \citep[e.g.,][]{Pushkarev2009, Linford2011, Pushkarev2012, Xiong2015}. Thus, the beamed emission from the jet dominates across multi-band radiations for these $\gamma$-ray emitting blazars. 

After correcting for the beaming effect and cosmological time dilation (or redshift) in blazars, the intrinsic optical variability timescale can be expressed as 
\begin{equation}
\tau^{\rm in}=\frac{\tau^{\rm obs}\times\delta}{1+z},
\end{equation}
where $\tau^{\rm obs}$ represents the observed timescale, $\delta$ is the Doppler factor, and $z$ is the redshift. Our Doppler factors were obtained by cross-matching with the sample of \citet{Liodakis2018}, who estimated the variability Doppler factor in the radio band for the largest collection of radio-bright blazars. The variability Doppler factor was derived from the relationship between variability brightness temperature and intrinsic brightness temperature.

Despite cross-matching, many blazars still lacked data on their variability Doppler factors. \citet{Nemmen2012} calculated the intrinsic $\gamma$-ray luminosity for blazars and gamma-ray bursts (GRBs) by correcting the observed $\gamma$-ray luminosity for the beaming factor $f_{\rm b}$, such that 
\begin{equation}
L^{\rm in}=f_{\rm b}L^{\rm obs},
\end{equation}
with $f_{\rm b}=1-{\rm cos}(1/\Gamma)$, where the jet opening angle satisfies $\theta < 1/\Gamma $ and $\Gamma$ denotes the bulk Lorentz factor. They utilized a power-law fit of $L^{\rm obs}$ versus $f_{\rm b}$ to estimate the beaming factor for blazars lacking direct measurements of $\delta$. 

Following a similar approach as \citet{Nemmen2012}, we fitted the relationship between $L^{\rm obs}$ and $f_{\rm b}$ from our sample, which was consistent with their findings. The power-law fit of $L^{\rm obs}$ versus $f_{\rm b}$ is presented in Fig. 3 and is expressed as 
\begin{equation}
f_{\rm b}=10^{-3.403}(\frac{L^{\rm obs}}{10^{49}})^{-0.3788\pm0.052}.
\end{equation}

The relationship between observed luminosity and intrinsic luminosity can also be expressed as 
\begin{equation}
L^{\rm in}=\frac{L^{\rm obs}}{\delta^{2+\alpha}},
\end{equation}
where $\alpha$ is the spectral index, defined as the photon spectral index minus 1 \citep{Urry1995}. Hence, we can derive 
\begin{equation}
{\rm log\delta}=\frac{{\rm log}f_{\rm b}}{-(2+\alpha)}.
\end{equation}

Thus, for blazars without measured variability Doppler factors, $\delta$ was evaluated based on the observed $\gamma$-ray luminosity and the power-law photon index. The $\gamma$-ray energy flux was K-corrected according to 
\begin{equation}
S_\gamma = S_\gamma^{\rm obs}(1 + z)^{\alpha - 1}.
\end{equation}

The $\gamma$-ray luminosity was then calculated using the relation 
\begin{equation}
L_\gamma = 4\pi d_L^2 S_\gamma,
\end{equation}
where $d_L$ is the luminosity distance.

The uncertainty in the variability Doppler factor was found to be $0.59$ (${\rm 1s.d.}/\sqrt{N}$), while the standard deviation of the residuals between the best-fit linear model and the data presented in Fig. 3 was $0.66$. All data points remained within the 2$\sigma$ prediction band after accounting for measurement errors. Consequently, the double standard deviation of the residuals was adopted as the uncertainty in $ {\rm log} f_{\rm b} $. Based on error propagation \citep[see][]{Nemmen2012}, the average uncertainty for $\delta$ for blazars lacking direct measurements was estimated to be $\approx 1.2$.

The physical properties of NLSY1 galaxies are similar to those of blazars \citep[e.g.,][]{Zhou2003, Foschini2011, Foschini2015}. The variability Doppler factors for NLSY1 galaxies from our sample, as reported by \citet{Liodakis2018}, were found to be greater than 5.0, necessitating a correction for the beaming effect. For the SSRQ, the variability Doppler factor was measured at 3.31, and a similar correction for the beaming effect was applied. Notably, even without applying the beaming correction for the SSRQ, our results remained consistent.

\begin{figure}
    \centering
    \includegraphics[width=0.45\textwidth]{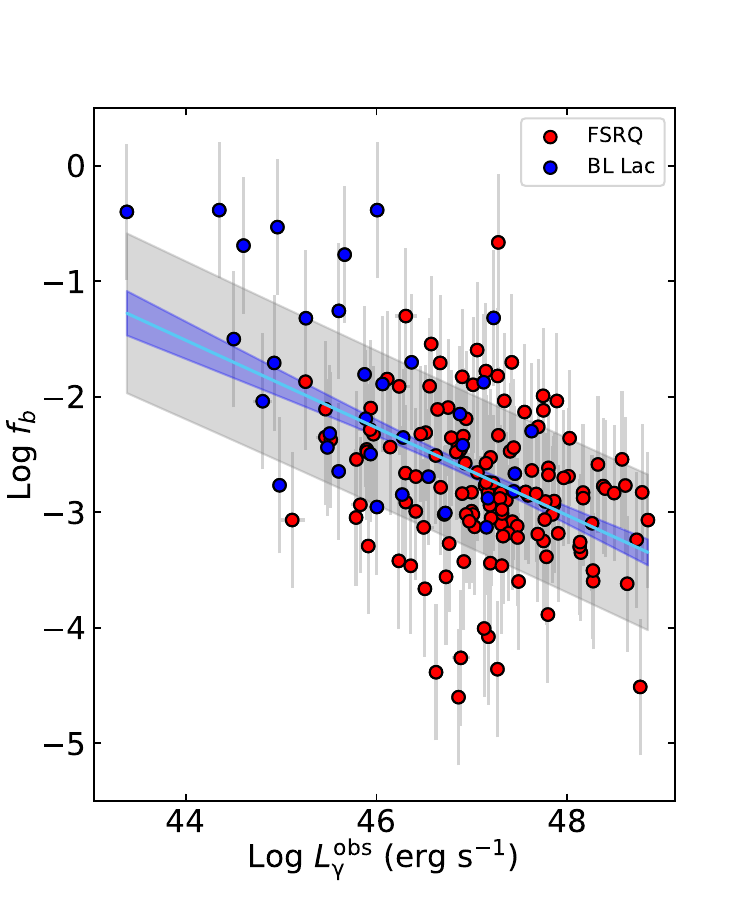}
    \caption{The relation between the observation $\gamma$-ray luminosity and the beaming factor. The blue solid line represents the best-fit linear model obtained through ordinary least squares fitting. The blue shaded region indicates the 1$\sigma$ confidence band, while the gray region depicts the 1$\sigma$ prediction bands. The correlation coefficient for this fit is $r = -0.49$, which corresponds to a confidence level of 6.8$\sigma$.}
\end{figure}

\begin{deluxetable*}{ccccccccccl}
\tablenum{3}
\tablewidth{0pt}
\tablecaption{The sample of jetted AGNs with inefficient accretion that exhibit good model fitting. The information on this table is the same as in the Table 2 except that $L_{\rm disk}$ is the upper limits. This table is available in its entirety in machine-readable form.\label{tab:3}} 
\tablehead{
\colhead{4FGL name} & $z$ &\colhead{Class} &\colhead{${\rm Log}~\tau$} &  \colhead{$\tau_{\rm le}$} & \colhead{$\tau_{\rm ue}$}  &  \colhead{${\rm Log}~M_{\rm BH}$} &\colhead{${\rm Log}~L_{\rm disk}$} & \colhead{${\rm Log}~L_{\rm \gamma}^{\rm obs}$}  &  \colhead{${\rm PL}_{\rm Index}$} &  \colhead{$\delta$}\\
\colhead{} & \colhead{} &  \colhead{} & \colhead{(days)}  &  \colhead{(days)} &\colhead{(days)} & \colhead{($M_{\odot}$)}  &  \colhead{(${\rm erg~s^{-1}}$)} &  \colhead{(${\rm erg~s^{-1}}$)}&  \colhead{}&  \colhead{}
}
\startdata
J0003.2+2207	&	0.1	&	bll	&	1.027	&	-0.246	&	0.273	&	8.1	$\pm$	0.11	&	42.74	&	43.24	$\pm$	0.11	&	2.12	$\pm$	0.18	&	2.47	$\pm$	1.22	 \\
J0006.4+0135	&	0.78	&	bll	&	1.622	&	-0.195	&	0.233	&	9.33	$\pm$	0.33	&	44.62	&	45.52	$\pm$	0.12	&	2.08	$\pm$	0.15	&	4.75	$\pm$	1.22	 \\
J0017.8+1455	&	0.3	&	bll	&	1.64	&	-0.129	&	0.162	&	8.27	$\pm$	0.44	&	44.16	&	44.98	$\pm$	0.06	&	2.25	$\pm$	0.11	&	3.79	$\pm$	1.22	 \\
J0021.6-0855	&	0.64	&	bll	&	2.081	&	-0.218	&	0.36	&	8.54	$\pm$	0.45	&	44.63	&	45.59	$\pm$	0.09	&	2.21	$\pm$	0.14	&	4.55	$\pm$	1.22	 \\
J0022.0+0006	&	0.3	&	bll	&	1.894	&	-0.18	&	0.251	&	8.02	$\pm$	0.4	&	43.79	&	44.69	$\pm$	0.1	&	1.59	$\pm$	0.15	&	4.82	$\pm$	1.22	 \\
J0033.3-2040	&	0.07	&	bll	&	1.396	&	-0.199	&	0.243	&	8.56	$\pm$	0.11	&	42.93	&	43.08	$\pm$	0.13	&	1.7	$\pm$	0.17	&	2.69	$\pm$	1.22	 \\
J0040.4-2340	&	0.21	&	bll	&	1.57	&	-0.229	&	0.253	&	8.68	$\pm$	0.13	&	43.75	&	44.44	$\pm$	0.07	&	2.22	$\pm$	0.13	&	3.31	$\pm$	1.22	 \\
J0059.3-0152	&	0.14	&	bll	&	1.63	&	-0.143	&	0.187	&	8.63	$\pm$	0.13	&	43.52	&	44.25	$\pm$	0.06	&	1.77	$\pm$	0.1	&	3.79	$\pm$	1.22	 \\
J0103.8+1321	&	0.49	&	bll	&	1.988	&	-0.179	&	0.258	&	9.69	$\pm$	0.25	&	44.41	&	45.38	$\pm$	0.07	&	2.1	$\pm$	0.12	&	4.52	$\pm$	1.22	 \\
\enddata
\end{deluxetable*}

\subsection{Black hole mass and Eddington ratio} \label{subsec:BH}

We cross-matched the 606 jetted AGNs with the $Fermi$ blazar sample from \citet{Paliya2021} to obtain the black hole masses and Eddington ratios. When estimating the black hole masses, \citet{Paliya2021} categorized the sample into two groups: emission-line blazars (predominantly jetted AGNs with radiatively efficient accretion) and absorption-line blazars (mostly jetted AGNs with radiatively inefficient accretion). For the emission-line blazars, the black hole mass was estimated using the traditional virial method based on single-epoch optical spectra. In contrast, for the absorption-line blazars, the black hole mass was computed from the stellar velocity dispersion or the absolute magnitudes of the host galaxy bulge.

The Eddington ratio is defined as $L_{\rm bol}/L_{\rm Edd} $, where $L_{\rm bol} = L_{\rm disk} \approx 10L_{\rm BLR} $ and $ L_{\rm Edd} = 1.3 \times 10^{38}(M/M_\odot)~{\rm erg~s^{-1}} $. Three sources—1H 0323+342, SBS 0846+513, and PKS 1502+036—were classified as NLSY1 galaxies in the 4LAC-DR3 catalog and in the sample of \citet{Foschini2022}. We maintained this classification, which differs from the categorization by \citet{Paliya2021}, which labeled them as blazars. Similarly, 3C 207 was classified as a SSRQ rather than a blazar.

The black hole masses for two FRI galaxies, PKS 0235+017 and B2 1113+29, were obtained by cross-matching with the $\gamma$-ray-emitting radio galaxies from \citet{Chen2023}. The black hole masses for these radio galaxies were estimated from the near-infrared magnitudes of their host galaxy bulges. Since there were no errors reported for the black hole mass estimates from \citet{Chen2023}, we estimated the errors for the two FRI galaxies using the average error of black hole masses in BL Lac objects. The uncertainty in the black hole masses from our sample reflects only the measurement uncertainty and does not account for systematic uncertainties.

The final sample of jetted AGNs with good model fitting, including redshift $z$, variability timescale  $\tau$, black hole mass $ M_{\rm BH} $, Eddington ratio $L_{\rm bol}/L_{\rm Edd} $ (or upper limits), observed $\gamma$-ray luminosity $L_{\rm \gamma}$, $\gamma$-ray photon index $PL_{\rm index}$, and Doppler factor $ \delta$, is presented in Tables 2 and 3. For the analysis, the sample is divided into two categories: emission-line and absorption-line AGNs, to account for the different methods of estimating black hole mass and the potential differences in the origins of optical variability. The dividing line distinguishing efficient accretion from inefficient accretion AGNs is approximately $ L_{\rm bol}/L_{\rm Edd} \approx 5 \times 10^{-3} $ \citep{Sbarrato2012}. Consequently, the emission-line AGNs are associated with efficient accretion when considering uncertainties (see Fig. A3 in the Appendix). Conversely, the absorption-line AGNs are classified as inefficient accretion AGNs, as upper limits on $L_{\rm BLR}$ are reported for our absorption-line AGN sample, with most having $L_{\rm bol}/L_{\rm Edd} < 5 \times 10^{-3}$ (see Fig. A3 in the Appendix). Non-jetted AGNs were sourced from the final sample presented by \citet{Burke2021}.

\begin{figure}
    \centering
    \includegraphics[width=0.48\textwidth]{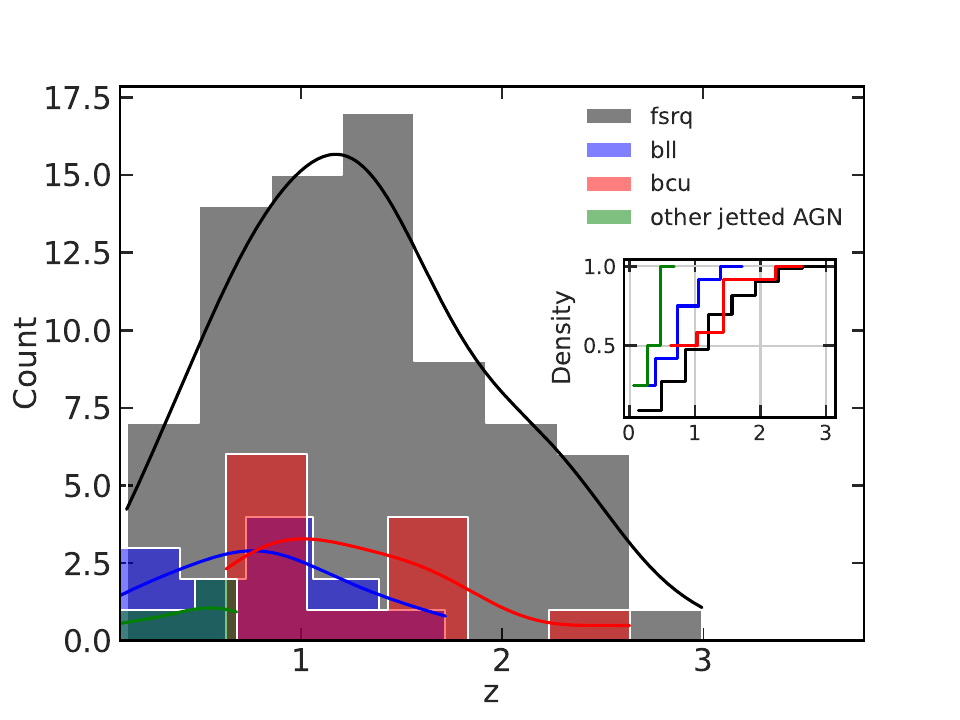}
    \includegraphics[width=0.48\textwidth]{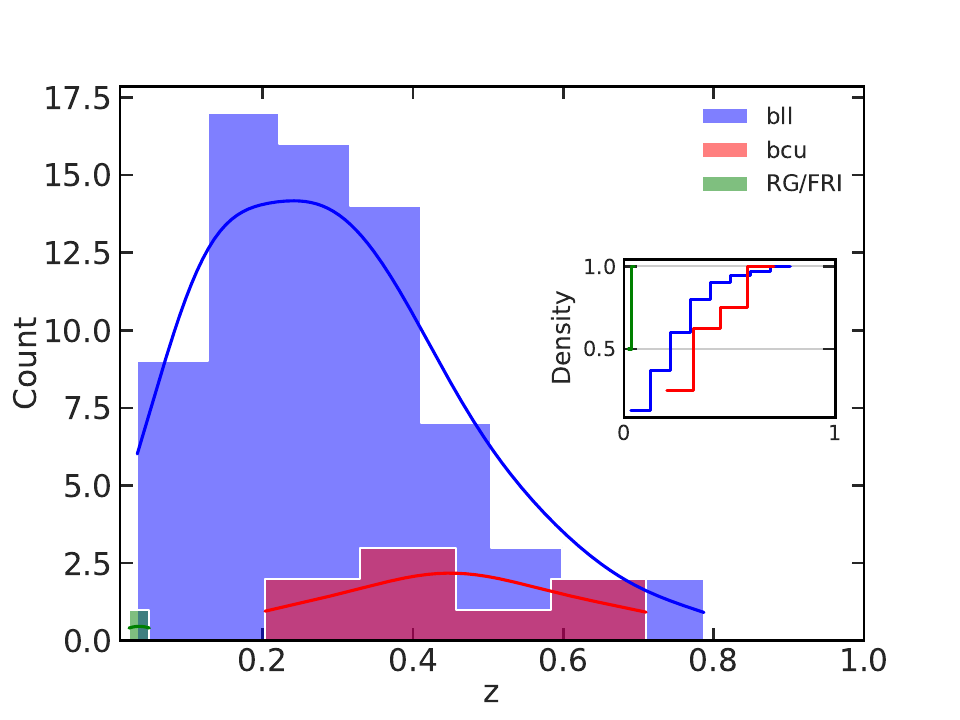}
    \caption{The histograms of redshift for jetted AGNs. The top panel displays jetted AGNs with efficient accretion, while the bottom panel shows those with inefficient accretion. The histograms, filled with different colors, represent the redshift distributions for various categories of jetted AGNs. Kernel density estimates are used to smooth the distributions, illustrated as lines of different colors. The sub-graph on the right represents the cumulative density, normalized so that the total area of the histogram equals 1. The \textit{bins} is set to \textit{`auto'}, which means that it is determined by the minimum bin width between the `Sturges' and `Freedman-Diaconis' estimator, offering good all-around performance. For specific details, please refer to the documentation for \textit{seaborn.histplot}.}
\end{figure}

\begin{figure}
    \centering
    \includegraphics[width=0.48\textwidth]{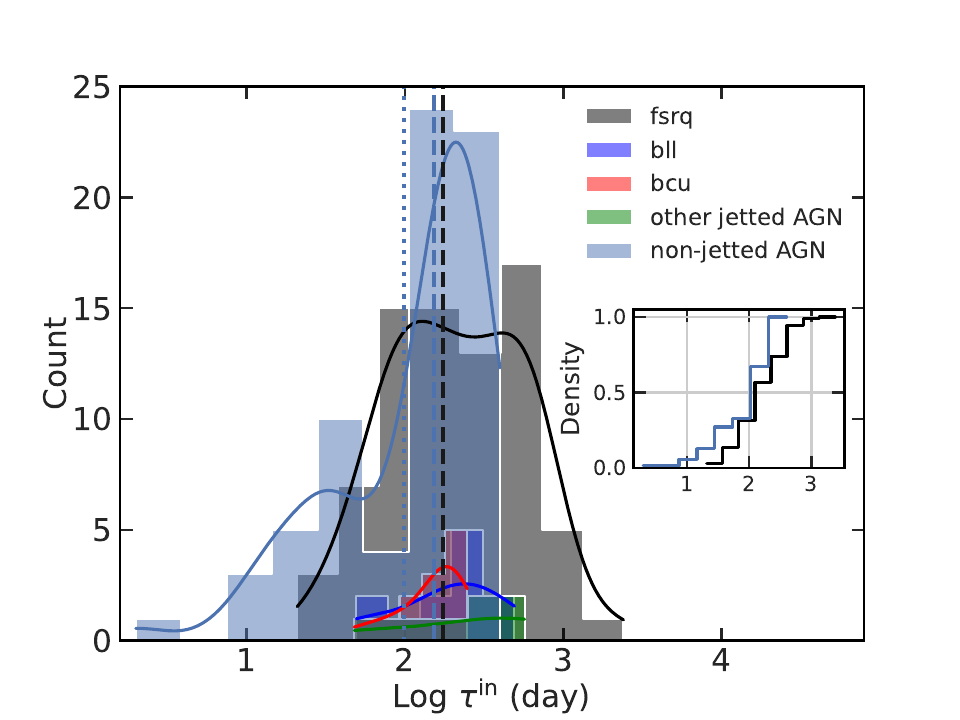}
    \includegraphics[width=0.48\textwidth]{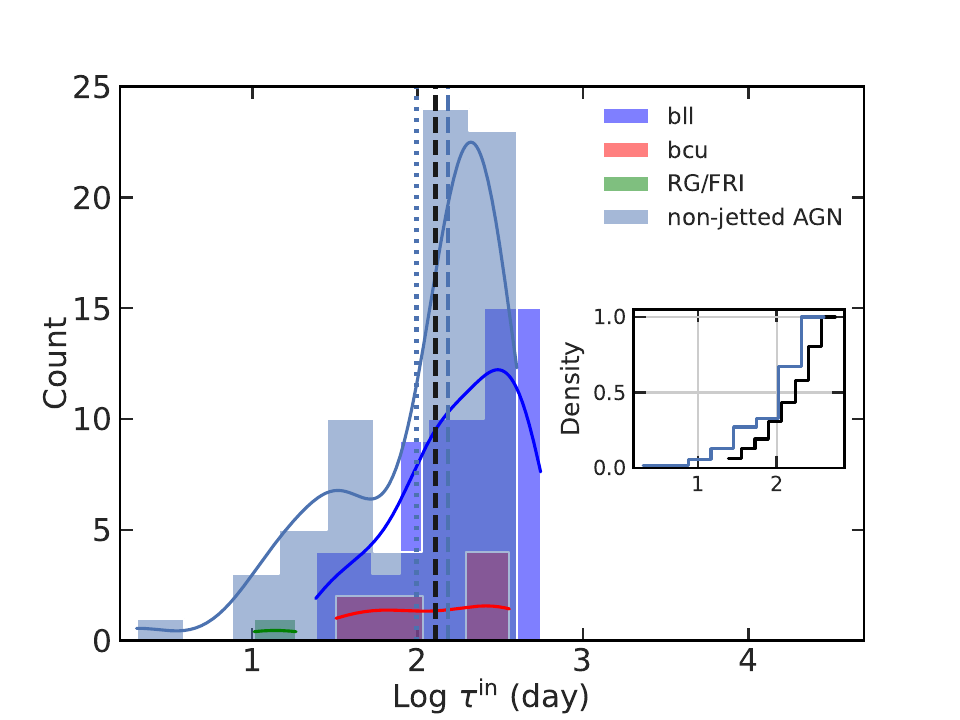}
    \caption{The histograms of intrinsic variability damping timescale. The top panel displays non-jetted AGNs and jetted AGNs with efficient accretion, while the bottom panel shows non-jetted AGNs and jetted AGNs with inefficient accretion. The histograms, filled with different colors, represent the variability timescale distributions for various categories of AGNs. Kernel density estimates are used to smooth the distributions, presented as lines of different colors. The sub-graph on the right shows the cumulative density, normalized so that the total area of the histogram equals 1. The black and blue lines in the sub-graph represent jetted AGNs and non-jetted AGNs, respectively. The dashed vertical lines indicate the average values of the timescale for jetted AGNs (black) and non-jetted AGNs (blue) with $M > 10^7 M_{\odot}$. The blue dotted vertical line represents the average timescale of non-jetted AGNs, including all black hole masses. The settings for the \textit{bins} are consistent with those used in Figure 4.}
\end{figure}

\begin{figure}
    \centering
    \includegraphics[width=0.62\textwidth]{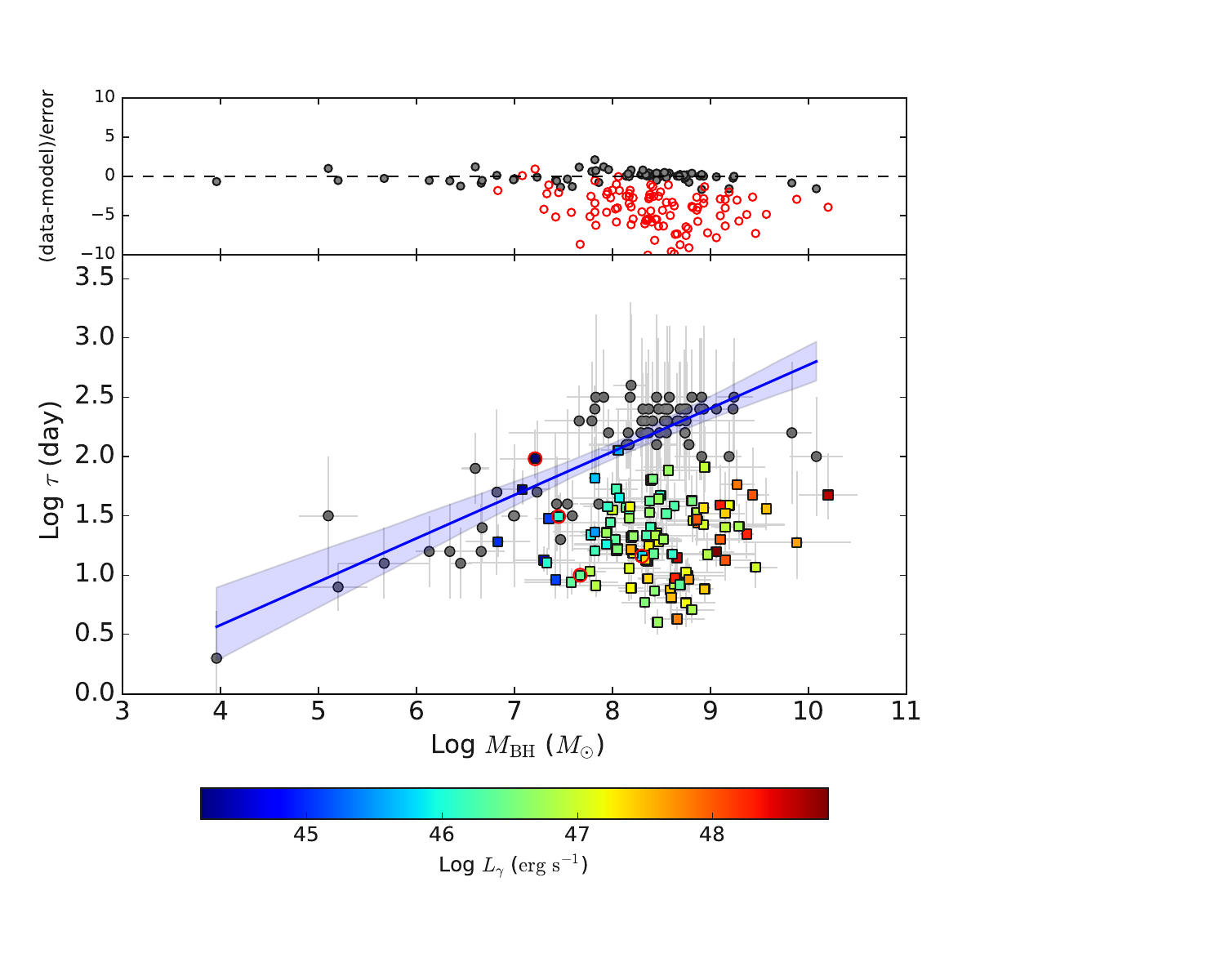}
    \includegraphics[width=0.62\textwidth]{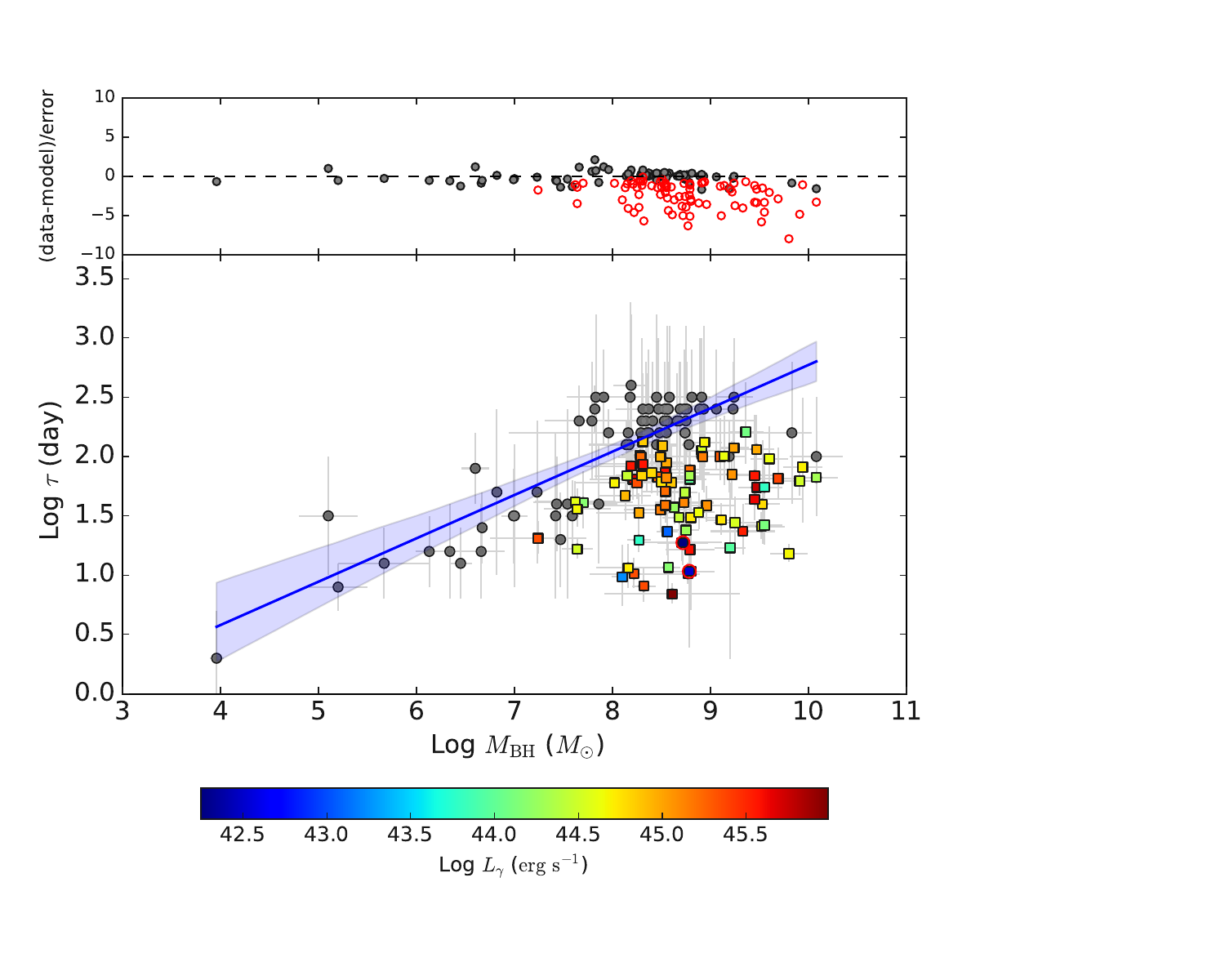}
    \caption{Optical variability damping timescale as a function of black hole mass for non-jetted and jetted AGNs after correcting for redshift. The top panel displays efficient accretion AGNs, while the bottom panel shows inefficient accretion AGNs. Grey solid circles represent non-jetted AGNs, and colored squares denote jetted AGNs, with squares featuring a red circle indicating non-blazar AGNs, while the others represent blazars. The color bars indicate $\gamma$-ray luminosity. The blue line represents the relationship defined by the best-fitting linear model for non-jetted AGNs, consistent with the findings of \citet{Burke2021}. The shaded interval represents the 95\% confidence bands. At the top of each panel, a sub-panel shows the residual plot after accounting for errors. In the residual plots, the red hollow circles represent jetted AGNs, and grey solid circles represent non-jetted AGNs.}
\end{figure}

\begin{figure}
    \centering
    \includegraphics[width=0.62\textwidth]{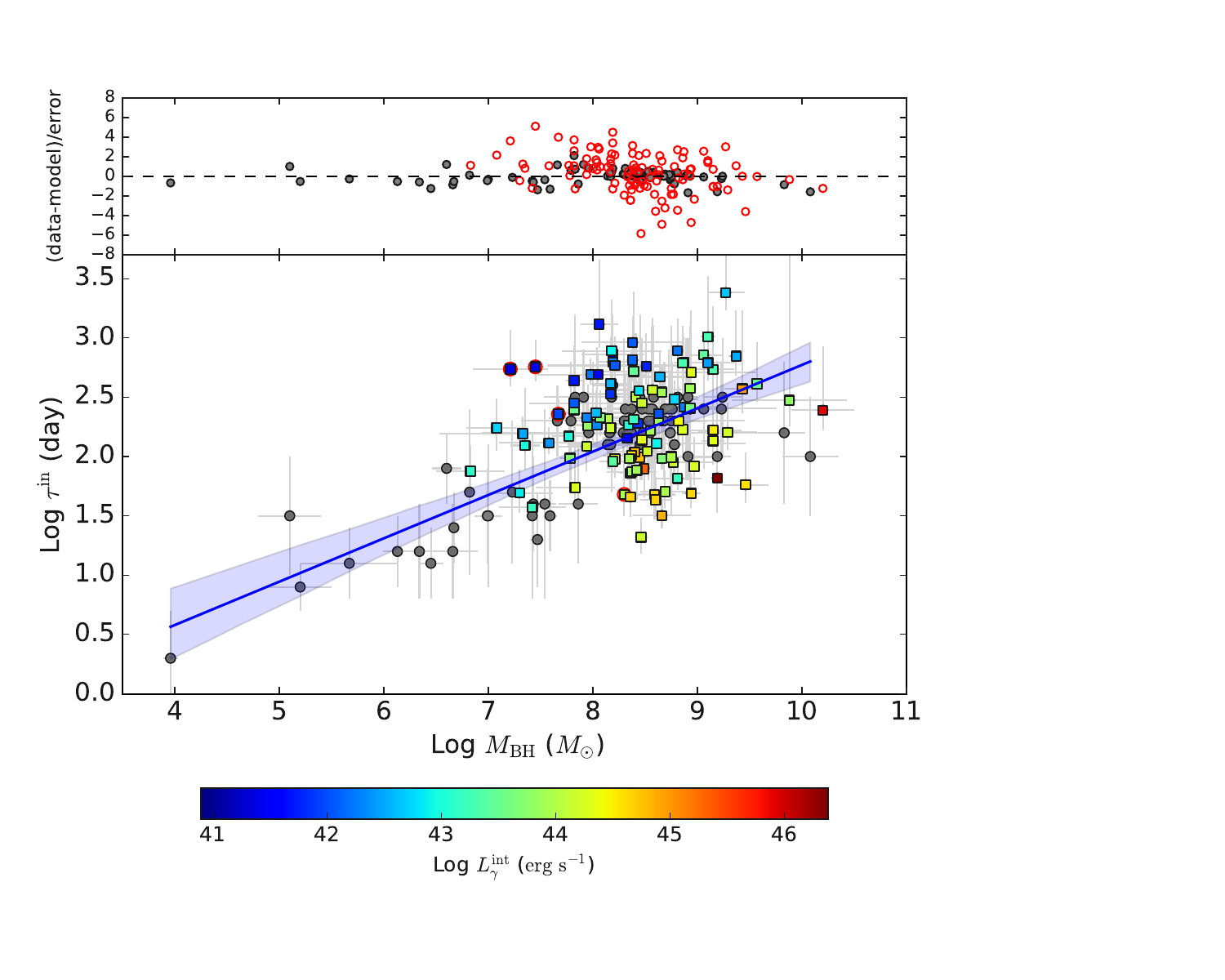}
    \includegraphics[width=0.62\textwidth]{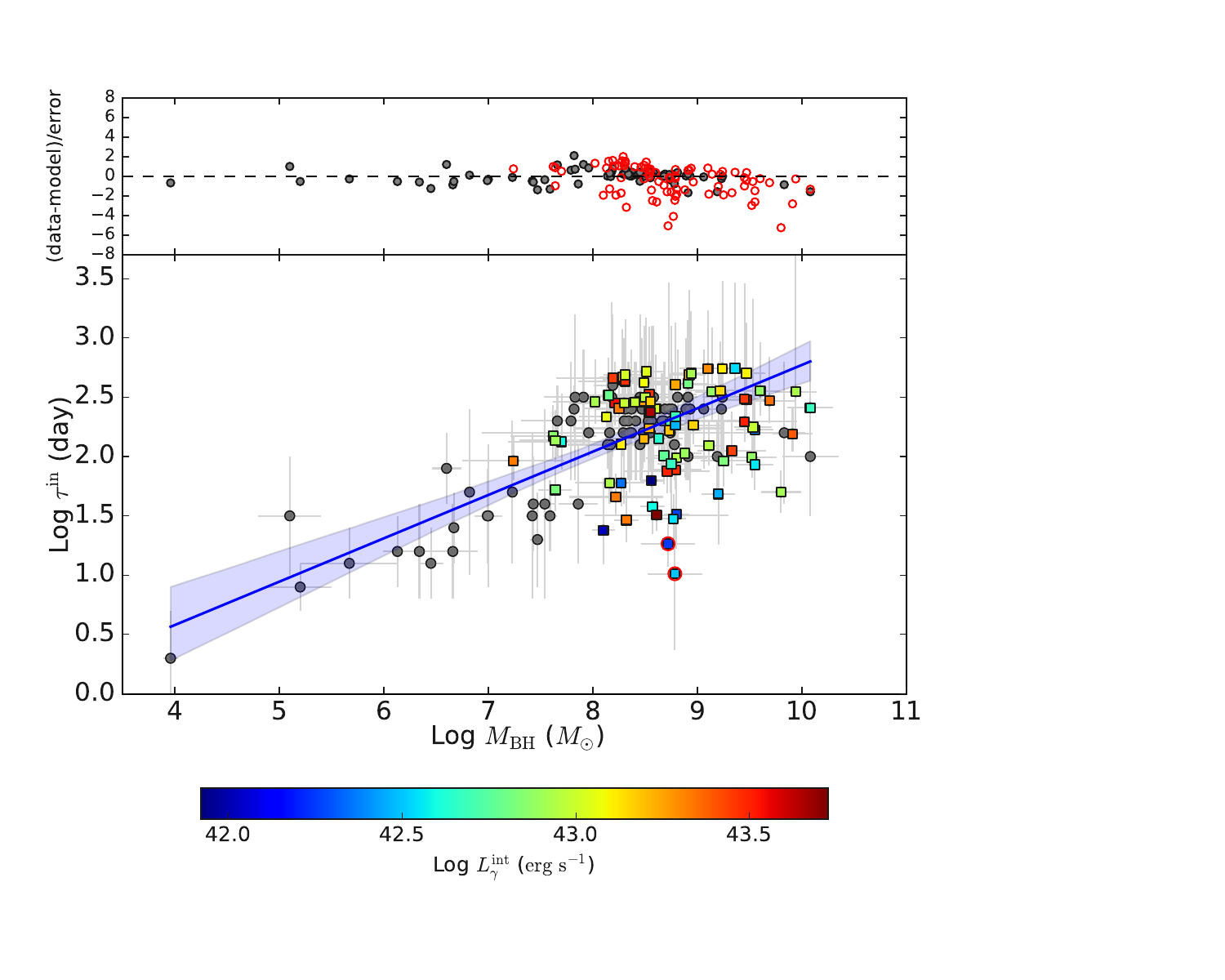}
    \caption{Intrinsic variability damping timescale as a function of black hole mass for non-jetted and jetted AGNs after correcting for the beaming effect and redshift. The color bars indicate intrinsic $\gamma$-ray luminosity. The other designations are the same as those in Fig. 6.}
\end{figure}

\begin{figure}
    \centering
    \includegraphics[width=0.5\textwidth]{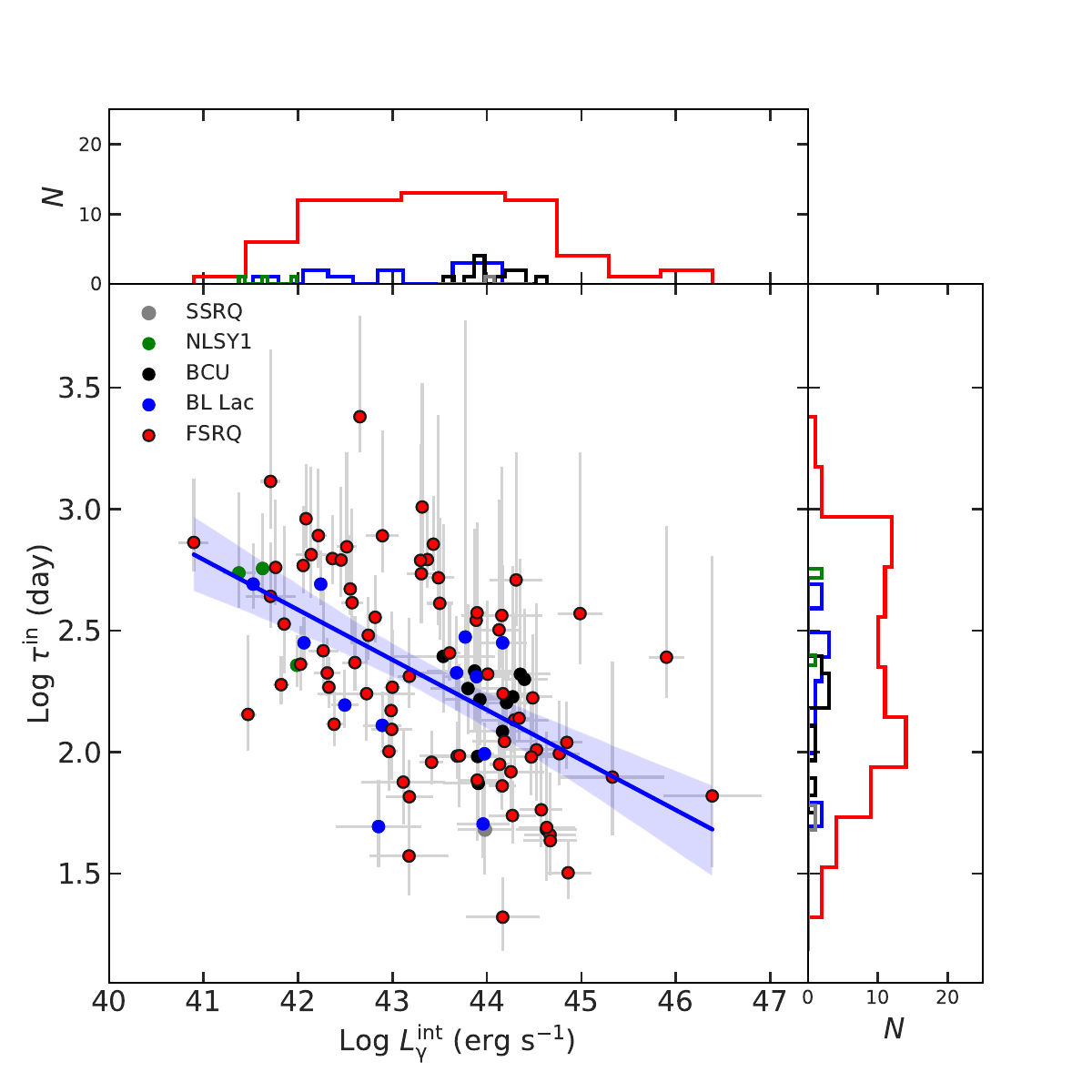}
    \includegraphics[width=0.5\textwidth]{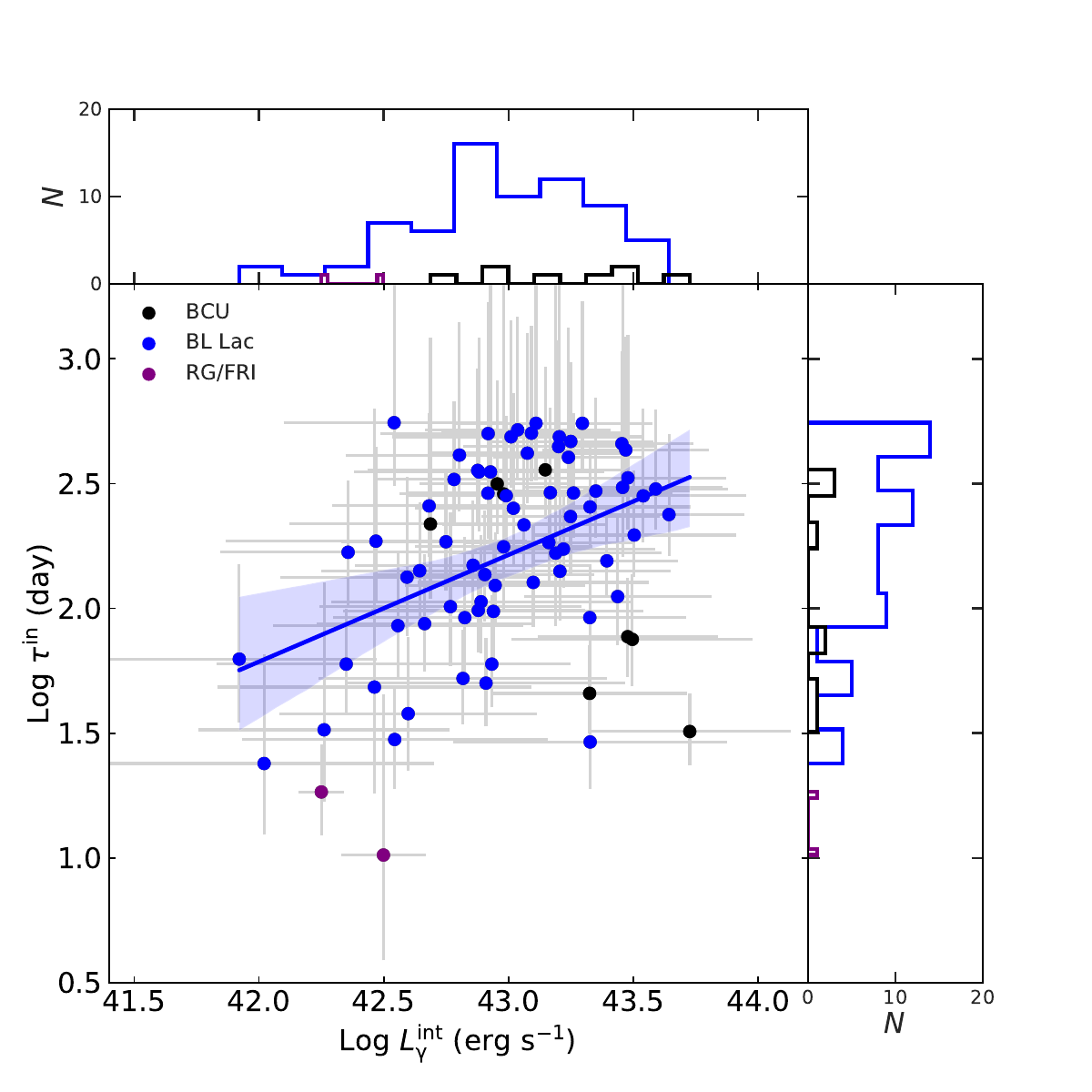}
    \caption{Intrinsic variability damping timescale as a function of intrinsic $\gamma$-ray luminosity for jetted AGNs after correcting for the beaming effect and redshift. The top panel displays efficient accretion AGNs, while the bottom panel shows inefficient accretion AGNs. The histograms of the parameters are provided at the top and right sides of each panel. The blue line represents the relationship defined by the best-fitting linear model, and the shaded interval indicates the 95\% confidence bands.}
\end{figure}

\begin{figure*}
\gridline{\fig{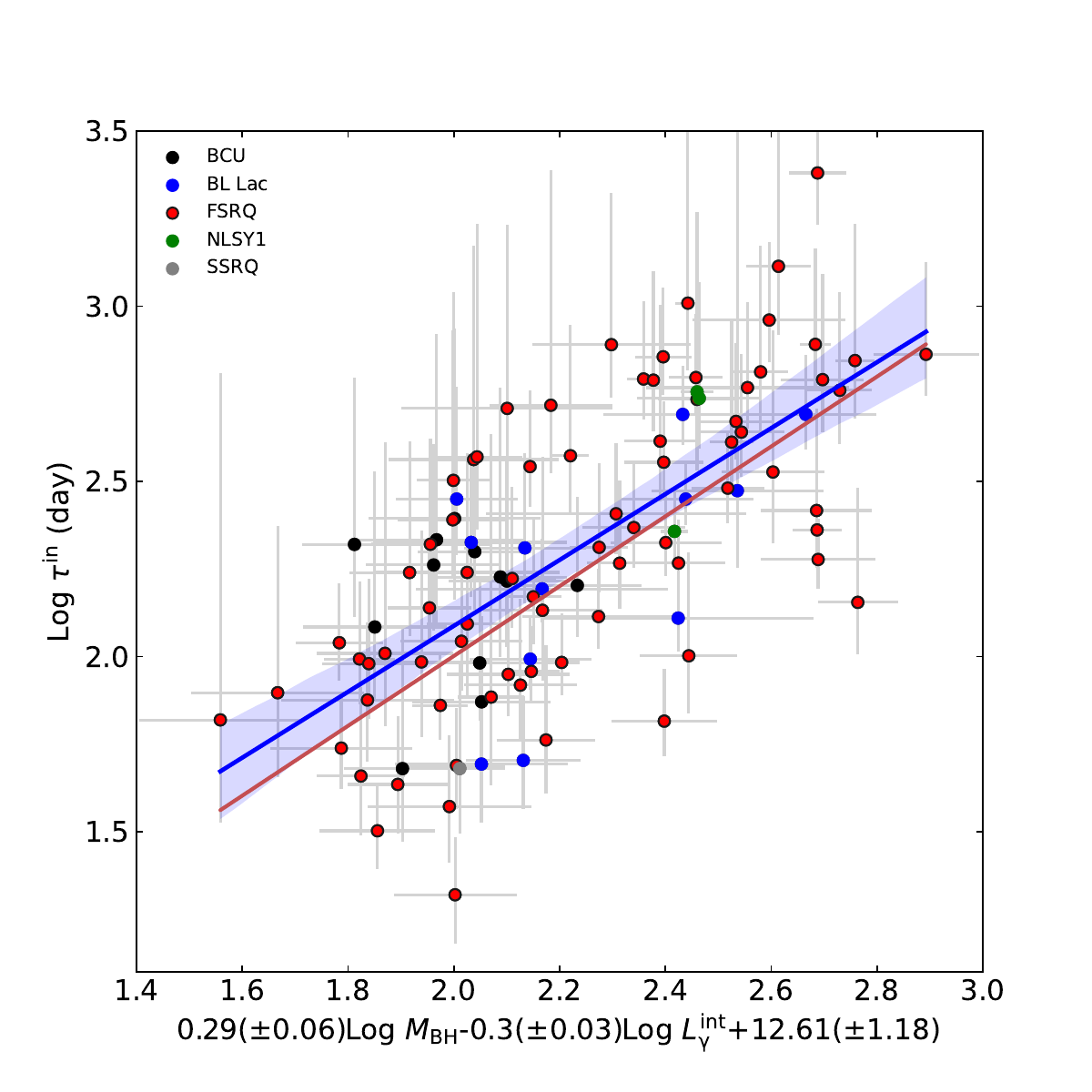}{0.5\textwidth}{(a) The efficient accretion AGNs with good fits of DRW model}
          \fig{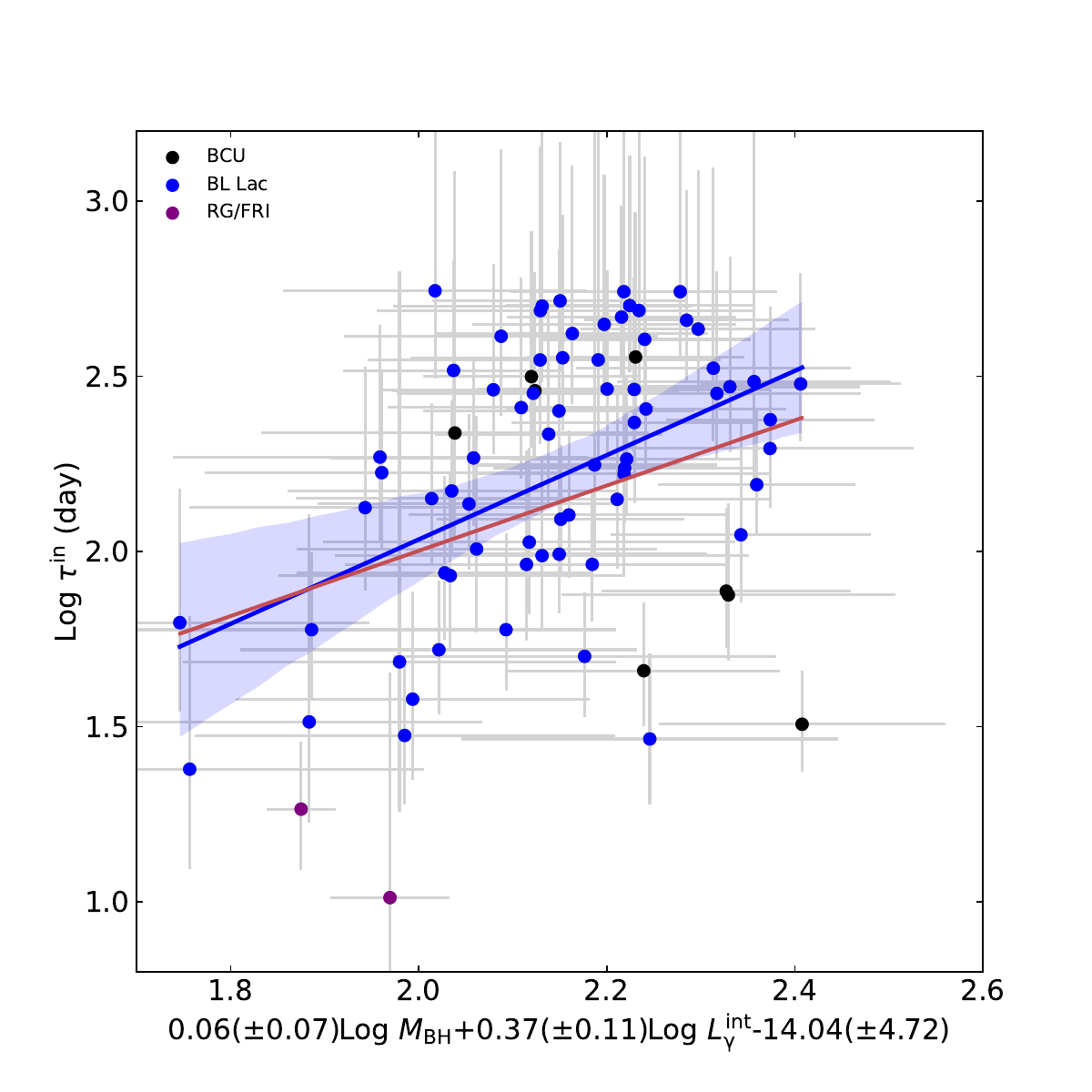}{0.5\textwidth}{(b) The inefficient accretion AGNs with good fits of DRW model}}
\gridline{\fig{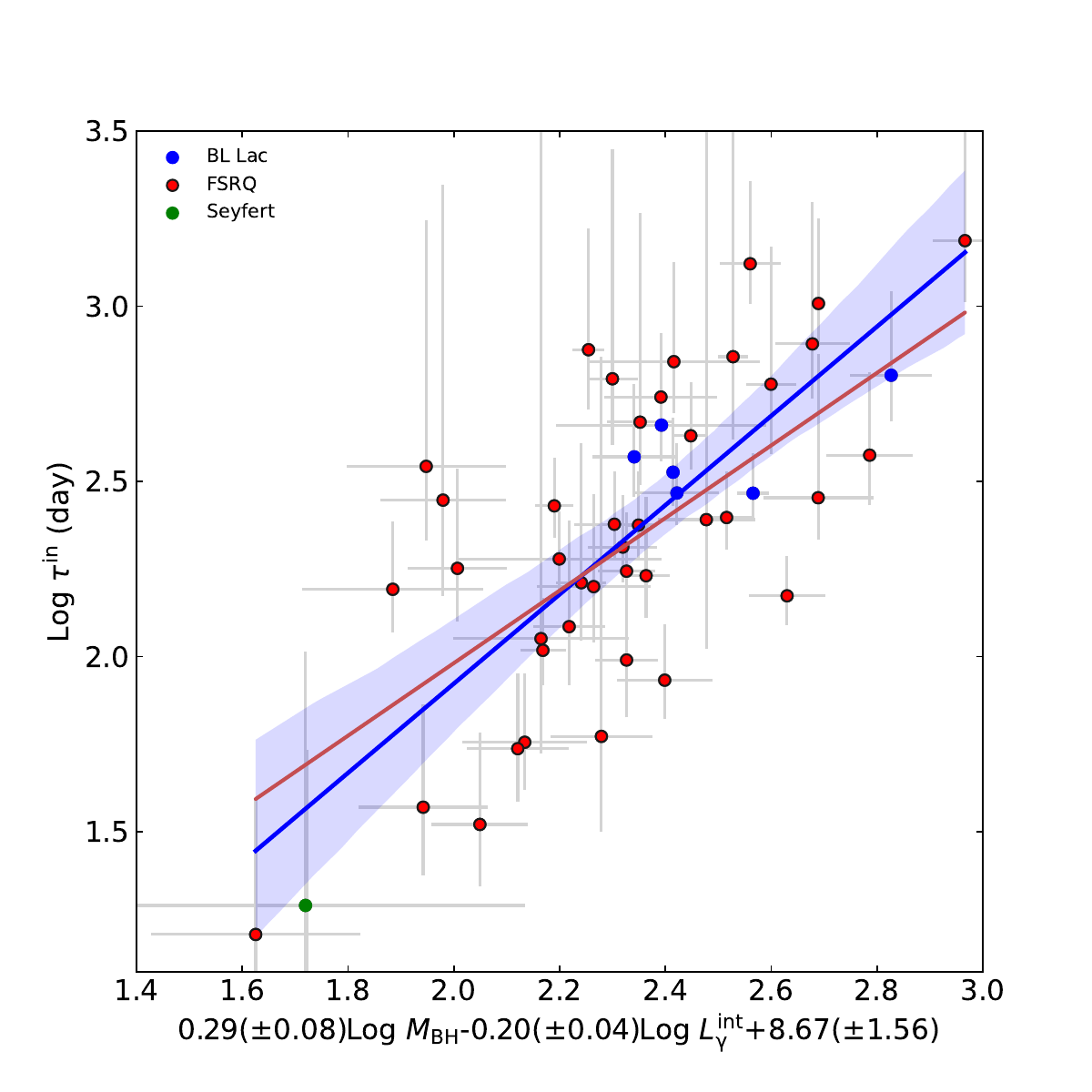}{0.5\textwidth}{(c) The efficient accretion AGNs with potentially good fits of DRW model}
          \fig{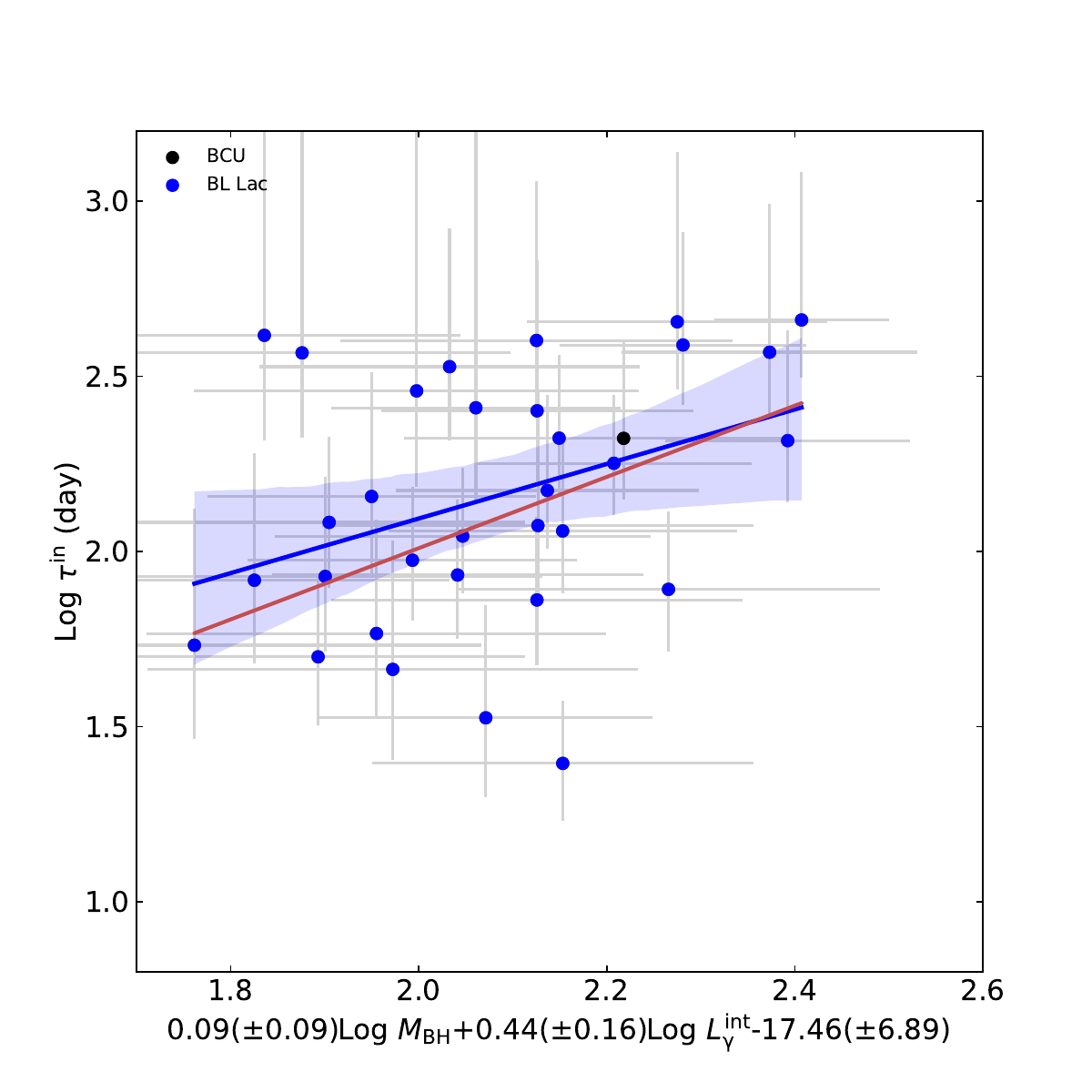}{0.5\textwidth}{(d) The inefficient accretion AGNs with potentially good fits of DRW model}}
    \caption{Intrinsic variability damping timescale vs. black hole mass and $\gamma$-ray luminosity for jetted AGNs. Panel (a) shows efficient accretion AGNs with good fits of the DRW model, while panel (b) shows inefficient accretion AGNs with good fits. Panels (c) and (d) correspond to the results for jetted AGNs with potentially good fits of the DRW model. The blue line represents multiple linear regression fitting for non-error weighted analysis, while the red line indicates error-weighted analysis. The shaded interval represents the 95\% confidence bands in the non-error weighted analysis.}
    \label{fig:multiple linear regression}
\end{figure*} 

\section{Results} \label{Results}
\subsection{The histograms of the parameters} \label{subsec:his}

The distribution range of redshift for efficient accretion AGNs is between 0 and 3, with an average value of 1.2 (see Fig. 4). In contrast, the redshift for inefficient accretion AGNs ranges from 0 to 0.8, with an average value of 0.3 (Fig. 4). The difference in redshift distribution between efficient and inefficient sources could result from selection bias, as efficient sources tend to be more luminous than inefficient ones.

The histograms of optical variability damping timescale, $\tau^{\rm in}$, are shown in Fig. 5. For efficient accretion AGNs, the range of $\tau^{\rm in}$ is from $10^{1.3}$ days to $10^{3.4}$ days, while for inefficient accretion AGNs, the range is from $10^{1.4}$ days to $10^{2.7}$ days. The corresponding error-weighted average values are $\langle \tau^{\rm in} \rangle = 10^{2.24 \pm 0.04}$ days for efficient accretion AGNs and $\langle \tau^{\rm in} \rangle = 10^{2.11 \pm 0.04}$ days for inefficient accretion AGNs. The error-weighted average of $\tau^{\rm in}$ for non-jetted AGNs is $10^{1.99 \pm 0.06}$ days, and $\langle \tau^{\rm in} \rangle = 10^{2.19 \pm 0.04}$ days for non-jetted AGNs with $M_{\rm BH} > 10^7 M_{\odot}$. Based on our results, $\langle \tau^{\rm in} \rangle$ ranges from weeks to years. The values of $\langle \tau^{\rm in} \rangle$ for jetted AGNs and non-jetted AGNs with $M_{\rm BH} > 10^7 M_{\odot}$ are nearly equal, while the average timescale for jetted AGNs with efficient accretion is approximately 45 days longer than that for jetted AGNs with inefficient accretion. 

\subsection{The relationships of correlation} \label{subsec:corr}

Most jetted AGNs and non-jetted AGNs are located in different regions on the panel of observed variability timescale versus black hole mass (Fig. 6). After correcting for the beaming effect, most jetted AGNs and non-jetted AGNs are now positioned in similar regions on the panel of intrinsic optical variability timescale versus black hole mass. The reasons are as follows: after correcting for the beaming effect, the average timescale value is more in line with that of non-jetted AGNs compared to the scenario where the beaming effect is not corrected. Additionally, after the beaming effect correction, the scatter around the best fit for non-jetted AGNs is smaller. Fig. 7 illustrates that jetted AGNs follow the relation defined by the best-fitting linear model from non-jetted AGNs ($\tau \propto M_{\rm BH}^{0.38}$). However, the scatter around the best fit for non-jetted AGNs is approximately 0.07 dex, while it is around 0.2 dex for jetted AGNs. When combining jetted AGNs and non-jetted AGNs into a single sample, we observe that the slope of the relationship between intrinsic optical variability timescale and black hole mass becomes flatter compared to that from non-jetted AGNs ($\tau \propto M_{\rm BH}^{0.3}$ for both efficient and inefficient accretion AGNs).

The residuals displayed in the top panel of each sub-graph in Fig. 7 represent the differences obtained by subtracting the best-fitting linear model for non-jetted AGNs from the observed data. These residuals are correlated with the intrinsic $\gamma$-ray luminosity for jetted AGNs (Fig. A4 in the Appendix).

The intrinsic $\gamma$-ray luminosity is used as a proxy for jet power, with the relationship given by ${\rm Log}P_{\rm jet} \propto {\rm Log}L^{\rm in}$ (where ${\rm Log}P_{\rm jet} = 0.98(\pm0.02){\rm Log}L^{\rm in} + 1.6(\pm0.9)$ as noted by \cite{Nemmen2012}). Fig. 8 demonstrates a trend of stronger jet power and shorter variability timescales for efficient accretion AGNs, while for inefficient accretion AGNs, the trend shows stronger jet power associated with longer variability timescales. A significant negative correlation between $L^{\rm in}$ and $\tau^{\rm in}$ is observed based on the results of the best linear fitting, with a correlation coefficient $r = -0.54$ and a significance level $P = 2 \times 10^{-9}$ where $P < 0.05$ indicates a significant correlation at the 95\% confidence level (see Fig. 8). To account for the negative correlation introduced by the beaming effect corrections (since $\tau^{\rm in} = \frac{\tau^{\rm obs} \times \delta}{1 + z}$ and $L^{\rm in} = \frac{L^{\rm obs}}{\delta^{2 + \alpha}}$), we conducted a partial correlation analysis to exclude the Doppler factor. The results of this analysis indicate that, even after removing the dependence on the Doppler factor, a significant negative correlation still exists between these variables ($P = 0.01$). For inefficient accretion AGNs, a significant positive correlation is found ($r = 0.4$ and $P = 1.9 \times 10^{-4}$; see Fig. 8).

Multiple linear regression fitting was employed to determine the relationship between $\tau^{\rm in}$, $M_{\rm BH}$, and $L^{\rm in}_{\gamma}$ for efficient accretion AGNs, yielding a correlation of $r = 0.68$ and $P = 3.6 \times 10^{-15}$. This significant correlation persists even after removing the dependence on the Doppler factor ($P = 7.3 \times 10^{-5}$). For error-weighted fitting, the weights are defined as $1/{\rm err}^2$, where ${\rm err} = \sqrt{(a \times X{\rm err})^2 + Y_{\rm err}^2}$, with $X_{\rm err}$ and $Y_{\rm err}$ being the errors on the x-axis and y-axis respectively, and $a$ being the non-error weighted fitting coefficient. Based on error-weighted fitting results and the relationship ${\rm Log}P_{\rm jet} \propto {\rm Log}L^{\rm in}$, the correlated relationship can be expressed as follows ($r = 0.7$, $P = 2.1 \times 10^{-15}$, and a scatter of approximately $0.09$ dex; Fig. 9a):
\begin{equation}      
\begin{split} 
{\rm Log}~\tau^{\rm in}=0.29(\pm0.06){\rm Log}~M_{\rm BH}-0.3(\pm0.03){\rm Log}~P_{\rm jet}\\+12.61(\pm1.18).  
\end{split}
\end{equation}

This indicates that $\tau^{\rm in}$ has a significant correlation with both $M_{\rm BH}$ and $P_{\rm jet}$, reflecting that both $M_{\rm BH}$ and $P_{\rm jet}$ contribute similarly to $\tau^{\rm in}$. When the Eddington ratio is also included, the correlated relationship is expressed as follows ($r=0.7$, $P=1.5\times10^{-14}$, and scatter of approximately 0.09 dex):
\begin{equation}      
\begin{split} 
{\rm Log}~\tau^{\rm in}=0.3(\pm0.06){\rm Log}~M_{\rm BH}-0.3(\pm0.03){\rm Log}~P_{\rm jet}\\+0.02(\pm0.05){\rm Log}~L_{\rm bol}/L_{\rm Edd}+12.76(\pm1.22), 
\end{split}
\end{equation}
which implies that $\tau^{\rm in}$ and $L_{\rm bol}/L_{\rm Edd}$ are either unrelated or weakly related (also see Fig. A3 in Appendix).

For the inefficient accretion AGNs, the results of the error-weighted fitting yield ($r=0.37$, $P=3\times10^{-3}$ and scatter of approximately 0.14 dex; Fig. 9b):
\begin{equation}      
\begin{split} 
{\rm Log}~\tau^{\rm in}=0.06(\pm0.07){\rm Log}~M_{\rm BH}+0.37(\pm0.11){\rm Log}~P_{\rm jet}\\-14.04(\pm4.72),
\end{split}
\end{equation}
indicating that $\tau^{\rm in}$ is primarily related to $P_{\rm jet}$ rather than $M_{\rm BH}$. It should be noted that the black hole mass spans six orders of magnitude in Fig. 7, whereas it spans about two orders of magnitude in this analysis.

For jetted AGNs with potentially good fits of DRW model, the results from the multiple linear regression fitting are consistent with the findings above, taking fitting errors into account (see Fig. 9c, 9d).

\section{Discussion} \label{discussion}
\subsection{DRW model fitting and parameter estimation} \label{subsec:Pe_DRW}

After applying a series of strict selection criteria, a substantial sample of gamma-ray emitting sources with long-term optical light curves was obtained. This ensures that the model fitting results are minimally affected by observational data, providing an opportunity to study the optical timescale separately for jetted AGNs with efficient and inefficient accretion.

Following the use of GPs to fit the DRW model to the long-term optical light curves, approximately 75\% of the sample demonstrates good or potentially good model fits, while around 25\% has poor fits. This indicates that a single DRW model from GPs can adequately fit the optical light curves of most jetted AGNs, although some jetted AGNs' light curves cannot be well-fitted by a single DRW model. Previous researches have shown that the optical light curves of some AGNs do not align well with this model \citep[e.g.,][]{Mushotzky2011, Smith2018, Caplar2017, Ding2024}. \citet{Ryan2019} found that the DRW model fails to capture the variability characteristics of blazars adequately, while higher-order CARMA models provide a more accurate description of the variability. For those jetted AGNs that are not well-fitted by the single DRW model, employing multiple DRW component models or alternative models may be necessary, along with a longer baseline and improved sampling.

For blazars lacking a radio Doppler factor, the 2$\sigma$ of the residual was selected as the uncertainty of ${\rm log}f_{\rm b}$ (i.e., the Doppler factor) to avoid underestimating uncertainty. The uncertainties in $\tau^{\rm in}$ and $L^{\rm in}$ primarily arise from the uncertainty propagation related to the Doppler factor. The black hole mass calculation considers only observational errors and does not account for systematic errors. If these blazars share similar systematic errors in black hole mass estimation, disregarding these systematic errors will not significantly impact the final correlation results obtained. 

\subsection{The distributions of optical variability damping time scales} \label{subsec:Op_v}

Previous studies have indicated that an insufficiently long baseline can lead to an underestimation of the fitted $\tau_{\rm DRW}$ \citep[e.g.,][]{Kozlowski2017, Kozlowski2021, Burke2021, Hu2024ApJ...961....5H}. Following the criteria established by \citet{Burke2021}, we selected observed $\tau^{\rm obs}_{\rm DRW}$ values less than $0.1 \times \text{baseline}$ (i.e., $<10^{2.28}$ days). It is important to note that some researchers have suggested that the baseline should exceed 10 or even 30 times the intrinsic $\tau_{\rm DRW}$ derived from theoretical models \citep[e.g.,][]{Kozlowski2017, Kozlowski2021, Zhou2024, Ren2024}. However, accurately determining the theoretical intrinsic timescale remains challenging due to the complex variability mechanisms present in jetted AGNs.

\citet{Zhang2022, Zhang2023} analyzed $\gamma$-ray, X-ray, and optical variability damping timescales from bright $\gamma$-ray emitting blazars using the celerite method, obtaining average timescales of several tens of days for these three bands after correcting for redshift. Our error-weighted average optical timescale is $10^{1.33 \pm 0.04}$ for jetted AGNs, which is generally consistent with the findings of \citet{Zhang2023}. The slight discrepancy, where our average timescale is lower than that reported by \citet{Zhang2023}, can be attributed to our larger sample, which includes jetted AGNs with smaller black hole masses and lower luminosities, as well as differences in the observational periods.

After correcting for the beaming effect, \citet{Ruan2012} estimated that the characteristic timescale of blazar variability in the optical band is approximately 3 years in the rest frame of the jet, assuming all blazars share the same Doppler factor of 10. Our intrinsic timescale is shorter than that of \citet{Ruan2012} by more than a factor of 6. A possible reason for this discrepancy is that the sample analyzed by \citet{Ruan2012} includes non-$\gamma$-ray-emitting blazars and encompasses different observation periods. Additionally, our criteria ensure that the observed timescale is less than $0.1 \times \text{baseline}$, which results in the longest observation timescale not exceeding $10^{2.28}$ days.

We find that the average intrinsic timescales $\langle\tau^{\rm in}\rangle$ for jetted and non-jetted AGNs with $M_{\rm BH}>10^7 M_{\odot}$ are nearly equal, implying that the variability timescale in jetted AGNs is related to the variability of the accretion disk, similar to non-jetted AGNs. Our results indicate that there are slight differences in average timescales between efficient and inefficient accretion AGNs, with efficient accretion AGNs exhibiting longer timescales. Many models predict such a difference \citep{Ryan2019}. Efficient accretion AGNs are expected to have standard Shakura-Sunyaev accretion disks \citep[SSAD;][]{Shakura1973}, while inefficient accretion AGNs likely operate through different mechanisms, such as advection-dominated accretion flows \citep[ADAF;][]{Narayan1997}. The characteristic fluctuation timescale in advection-dominated disks is expected to be shorter than that in SSAD, according to the fluctuating $\alpha$ model proposed by \citet{Lyubarskii1997}.

\subsection{The correlations between variability damping time scale and other physical parameters} \label{subsec:Co_b}

For non-jetted AGNs, numerous studies have identified correlations between optical damping timescales and various physical measurements, including black hole mass, monochromatic luminosity, absolute magnitude, wavelength, and accretion rate \citep[e.g.,][]{Kelly2009, MacLeod2010, Suberlak2021, Burke2021, Zhou2024, Ren2024, Su2024ApJ...969...78S}. \citet{Burke2021} found a correlation between timescale and black hole mass that extends across the entire mass range of supermassive black holes, with the relation given by $\tau \propto M_{\rm BH}^{0.38}$ for non-jetted AGNs. In the framework of standard accretion disks, thermal timescales scale with mass as $\tau \propto M_{\rm BH}^{0.5}$. The dominant variability originates within the disk at the UV-emitting radii, and this variability is then communicated to other radii in the disk \citep{Burke2021}.

In our results, most jetted AGNs and non-jetted AGNs are located in the same regions of the panel depicting intrinsic variability timescale versus black hole mass after correcting for the beaming effect. These findings support the idea that the variability in both jetted and non-jetted AGNs shares a common origin, suggesting that the optical variability of jetted AGNs may also originate within the accretion disk at UV-emitting radii. However, when jetted AGNs and non-jetted AGNs are combined into a single sample, the slope of the relationship between intrinsic optical variability timescale and black hole mass becomes flatter compared to that in non-jetted AGNs, and the scatter associated with jetted AGNs is larger. This indicates that the variability timescale for jetted AGNs is influenced not only by changes in the accretion disk but also by additional factors. Moreover, it is essential to consider efficient and inefficient accretion AGNs separately, as they may involve different jet and accretion mechanisms.

\subsubsection{The jetted AGNs with efficient accretion}

These residuals, obtained by subtracting the best-fitting linear model for non-jetted AGNs from the data, are associated with the intrinsic $\gamma$-ray luminosity. This implies that the larger scatter observed in jetted AGNs is related to the intrinsic $\gamma$-ray luminosity, suggesting that the intrinsic variability timescale in jetted AGNs is related not only to black hole mass but also to intrinsic $\gamma$-ray luminosity or jet power.

To further explore the origin of variability in jetted AGNs, we incorporate jet power and accretion rate into the analysis. The $\gamma$-ray luminosity is chosen as a proxy for jet power. The dependence of damping timescales on rest wavelength is not examined, as we only analyze a single optical band. After introducing jet power as a new parameter, the coefficients for black hole mass and jet power in the multiple regression equation are nearly identical for efficient accretion AGNs, indicating that jet power is also related to intrinsic variability timescale, exhibiting a negative correlation between the two.

The flux enhancement caused by the Doppler factor is a natural explanation for the variability seen in beaming AGNs. If the Doppler factor remains relatively constant during the observation period, we have nearly completely corrected for the beaming effect in our results. Long-term variability in beaming AGNs may arise from changes in the orientation of the jet-emitting region. However, the assumption of a constant Doppler factor during the observation period could introduce scatter in the estimation of the intrinsic long-term timescale, particularly for timescales on the order of years. The deviations from the best fit observed in Fig. 9 may partly result from this assumption. For some AGNs with efficient accretion, if the radiation from the disk is significant compared to the beamed radiation from the jet, then the corrected damping timescale should be longer than the actual value. The upward scatter observed in Figure 9(a) may be due to beaming effect corrections, especially at longer timescales.

As jet power decreases, the radiation from the accretion disk becomes more prominent. If the characteristic timescale associated with disk radiation exceeds that of jet radiation, then lower jet power is accompanied by longer timescales, which explains the negative correlation between $P_{\rm jet}$ and $\tau^{\rm in}$. In this scenario, we would also expect a negative correlation between $\tau^{\rm in}$ and the dominant jet parameter ($F_{\rm j} = L^{\rm obs}_{\gamma}/L_{\rm disk}$), which represents the relative significance of beamed luminosity from the jet compared to disk radiation. However, this is not consistent with our analysis (see Fig. 10), indicating that such an explanation does not account for our results.

\begin{figure}
    \centering
    \includegraphics[width=0.45\textwidth]{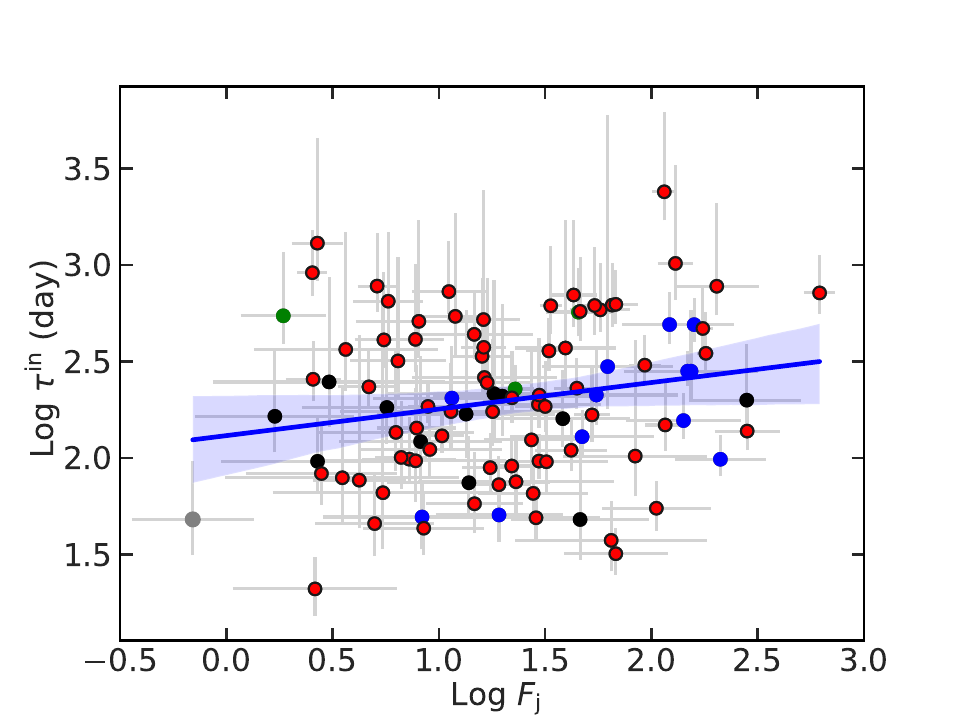}
    \caption{Intrinsic variability damping time scale as a function of the dominant parameter of jet. The meanings of the symbols are consistent with those used in Fig. 8.}
\end{figure}

Variations in physical parameters (e.g., magnetic field strength, particle acceleration rate, cooling rate, escape rate) can produce variability in the jet-emitting region (or blob). The characteristic time scales of variability in jetted AGNs correspond to these time scales (e.g., light-crossing timescale, cooling timescale, acceleration timescale, escape timescale). The intrinsic cooling timescale is expressed as \citep{Danforth2013,Ghisellini2013}:

 \begin{equation}
 t_{\rm cool} = \frac{3m_{\rm e}c}{4\sigma_{\rm T}u_0'\gamma_{\rm e}}=1.2~\rm yr \Big(\frac{10^3}{\gamma_{\rm e}}\Big)\Big(\frac{0.1~\rm G}{\it{B'}}\Big)^2\Big(\frac{2}{1+\it{q}}\Big)\,,
 \end{equation}
where $\gamma_{\rm e}$ is the characteristic random Lorentz factor of electrons, and $u_0' = u_B' + u_{\rm rad}' = \frac{(1+q)B'^2}{8\pi}$ (with $u_B'$ and $u_{\rm rad}'$ representing the co-moving energy densities of the magnetic field and soft photons, respectively). The Compton dominance parameter $q = \frac{u_{\rm rad}'}{u_B'} \simeq \frac{L_{\rm IC}}{L_{\rm syn}}$, $\sigma_{\rm T}$ is the Thomson scattering cross-section, and $m_{\rm e}$ is the mass of the electron \citep{Xiong2020}. 

The jet power is predicted to depend on $B'^2$ \citep[$P_{\rm jet} \propto B'^2$;][]{Blandford1977, Ghisellini2014}. Thus, the relationship between jet power and cooling timescale can be articulated as $t_{\rm cool} \propto P_{\rm jet}^{-1}$.

When particle acceleration is dominated by diffusive shock acceleration, the acceleration timescale in the relativistic limit is given by \citep{Protheroe2004, Rieger2007}:
 \begin{equation}
 \begin{split}
 t_{\rm acc} \simeq \frac{3\alpha r_{\rm L}}{20c}\simeq\frac{20\alpha \gamma_{\rm e}m_{\rm e}c}{3eB'}\\ =1.2~\rm yr \Big(\frac{\gamma_{\rm e}}{10^6}\Big)\Big(\frac{10^{-3}~\rm G}{\it{B'}}\Big)\Big(\frac{\alpha}{10^5}\Big)\,,
 \end{split}
 \end{equation}
under the quasi-linear theory, where $r_{\rm L}$ is the Larmor radius and $\alpha$ is the ratio of the mean magnetic field energy density to the turbulent magnetic field energy density \citep{Xue2019}. The relationship between jet power and acceleration timescale is expressed as $t_{\rm acc} \propto P_{\rm jet}^{-0.5}$. From Equations 14 and 15, we can infer that timescales on the order of years are possible under a relatively weak magnetic field configuration.

By comparing our results (Equations 11 and 12) with the above theories, we find that the intrinsic variability timescale from efficient accretion AGNs aligns more closely with the timescale of diffusive shock acceleration. The Eddington ratio does not significantly contribute to the damping timescale for efficient accretion AGNs, indicating that variations in accretion rate are not the primary source of optical variability.

Based on the preceding discussion, we propose the following physical scheme: For AGNs with efficient accretion, the optical variability of jetted AGNs may originate within the standard accretion disk at UV-emitting radii, similar to non-jetted AGNs. In this context, the accretion rate remains constant. Magnetic fields may play a crucial role throughout this process. Different radii within the accretion disk could be coupled through large-scale magnetic fields, with variability at UV-emitting radii potentially launching Alfve\'n waves into optical-emitting radii, thereby driving dissipative heating at similar variability timescales to those at the launching radius \citep{Burke2021}. The variability originating from the accretion disk can trigger changes in the jet, potentially through the magnetic field because of the coupling between the jet and the disk. The shock generated at the base of the jet may be influenced by this magnetic field. As the shock propagates along the jet, it sweeps through the radiation zone. In this radiation region, electrons are primarily accelerated by the diffusive shock, resulting in the observed optical variability. This variability is further enhanced by the beaming effect in beamed AGNs. However, it is noted that in addition to the aforementioned explanation, there may be other potential mechanisms underlying the origin of variability.

\subsubsection{The jetted AGNs with inefficient accretion}

The escape times scale of diffusion can be expressed as \citep{Xue2019}:
 \begin{equation}
 \begin{split}
 t_{\rm esc}=\frac{R^2}{4D} = \frac{3eB'R^2}{4\alpha \gamma_{\rm e}m_{\rm e}c^3}\\=1.2~\rm yr \Big(\frac{10^6}{\gamma_{\rm e}}\Big)\Big(\frac{\it{B'}}{10^{-3}~\rm G}\Big)\Big(\frac{10^5}{\alpha}\Big)\Big(\frac{\it{R}}{5\times10^{17}~\rm cm}\Big)^2\,,
 \end{split}
 \end{equation}
where $D$ is the diffusion coefficient and $R$ is the radius of the emission zone. The relationship between jet power and escape time scale is given by $t_{\rm esc}\propto P_{\rm jet}^{0.5}$. For inefficient accretion AGNs, the intrinsic variability timescale aligns with the escape timescale of diffusion when comparing our results (see Equation 13) with the aforementioned theories. 

It is important to note that when considering the correlation coefficient $r$, the correlation for inefficient accretion AGNs is moderate (see Equation 13). The brightness is lower in AGNs with inefficient accretion, which results in greater uncertainty in the damping timescale for most of these AGNs compared to those with efficient accretion. Uncertainties arising from the Doppler factor and damping timescale may cause the data to deviate from the best-fit model, contributing to this moderate correlation.

Based on the above discussion, we propose the following physical scheme: For jetted AGNs with inefficient accretion, variability may arise from changes within the inefficient accretion disk or the jet itself. Once shock or turbulence is generated, particles are accelerated, undergo radiation cooling, and escape, with the escape process serving as the dominant factor. Variability in jet results from changes in the escape rate, which is further enhanced by the beaming effect in beamed AGNs. 

\section{Conclusion}  \label{conclusion}
The damped random walk model, implemented through Gaussian Processes, was used to fit the $ZTF$ long-term optical light curves of 1684 $\gamma$-ray-emitting jetted AGNs. 
This represents one of the largest samples to date with characteristic optical variability timescales for jetted AGNs (Table 1).
Approximately $75\%$ of the sample shows good or potentially good model fits, while the remaining $25\%$ exhibits poor fits. 
While a single DRW model adequately describes the optical variability of most jetted AGNs, some AGNs deviate, indicating the need for alternative models for certain cases.
For these jetted AGNs with well-fitted DRW models, three additional criteria were applied to ensure reliable measurements of the damping timescale. The final jetted AGNs sample, with detailed parameters including $z$, $\tau$, $M_{\rm BH}$, $L_{\rm bol}/L_{\rm Edd}$ (or upper limit), $L_{\gamma}$, spectral index, and $\delta$, is given in Table 2 and  3. 

This large and diverse sample enables a detailed analysis of variability properties across efficient and inefficient accretion regimes. The main conclusions are as follows.

(i) For $\gamma$-ray emitting jetted AGNs with efficient and inefficient accretion, the intrinsic variability timescale $\tau^{\rm in}$ ranges from $10^{1.3}$ days to $10^{3.4}$ days and $10^{1.4}$ days to $10^{2.7}$ days, respectively. The corresponding error-weighted average value is $\langle \tau^{\rm in} \rangle = 10^{2.24 \pm 0.04}$ days for accretion efficient AGNs, slightly larger than that for accretion inefficient AGNs ($\langle \tau^{\rm in} \rangle = 10^{2.11 \pm 0.04}$ days).  

(ii) Most jetted AGNs and non-jetted AGNs are found in the same regions on the panel depicting intrinsic variability timescale versus black hole mass after correcting for the beaming effect. The jetted AGNs tend to follow the relation defined by the best-fitting linear model established for non-jetted AGNs. These results suggest that both jetted and non-jetted AGNs share a common origin of variability. 

However, when jetted AGNs and non-jetted AGNs are combined into a single sample, the slope of the relationship between intrinsic variability timescale and black hole mass becomes flatter compared to that observed for non-jetted AGNs, and the scatter for jetted AGNs is larger. This indicates that the variability timescale in jetted AGNs is influenced not only by changes in the accretion disk but also by other factors.

(iii) In addition to the $M_{\rm BH}$, the $P_{\rm jet}$, proxied by $\gamma$-ray luminosity, is introduced as a new parameter to explore the correlations between variability damping time scale and other physical parameters. New relationships among $\tau^{\rm in}$, $M_{\rm BH}$ and $P_{\rm jet}$, obtained by multiple linear regression fitting, are discovered for efficient accretion AGNs ($\tau^{\rm in} \propto M_{\rm BH}^{0.29^{+0.06}_{-0.06}}P_{\rm jet}^{-0.3^{+0.03}_{-0.03}}$ with scatter of approximately 0.09~dex) and for inefficient accretion AGNs ($\tau^{\rm in} \propto M_{\rm BH}^{0.06^{+0.07}_{-0.07}}P_{\rm jet}^{0.37^{+0.11}_{-0.11}}$ with scatter of approximately 0.14~dex), respectively. 

For efficient accretion jetted AGNs, a significant negative correlation exists between $\tau^{\rm in}$ and $P_{\rm jet}$, with $M_{\rm BH}$ and $P_{\rm jet}$ contributing nearly equally to $\tau^{\rm in}$. The agreement between observations and theoretical predictions provides new evidence supporting the notion that variability is related to the acceleration of diffusive shocks.

In contrast, for inefficient accretion jetted AGNs, $M_{\rm BH}$ does not significantly contribute to $\tau^{\rm in}$. Instead, $\tau^{\rm in}$ is primarily (and positively) related to $P_{\rm jet}$. The intrinsic timescale is consistent with the escape timescale of electrons, as evidenced by a comparison of our results with the theoretical relationship between $\tau^{\rm in}$ and $P_{\rm jet}$.

(iv) Our results suggest that the optical variability of jetted AGNs with efficient accretion may originate within the standard accretion disk at UV-emitting radii, similar to non-jetted AGNs. In the radiation region of the jet, electrons are primarily accelerated by diffusive shocks, leading to the observed optical variability. This variability is further enhanced by the beaming effect in beamed AGNs. Magnetic fields may play a significant role throughout this process.

For jetted AGNs with inefficient accretion, variability may arise from changes in the inefficient accretion disk or the jet itself. Variability in jet is related to variations in the escape rate and is also enhanced by the beaming effect in beamed AGNs.

\vspace{5mm}
\noindent 
We sincerely thank the referee for valuable comments and suggestions. 
This work is supported by the National Key R\&D Program (2023YFE0101200), the National Natural Science Foundation of China (Grant Nos. 12473020 and 12393813), the Yunnan Province Youth Top Talent Project (Grant No. YNWR-QNBJ-2020-116), the Yunnan Revitalization Talent Support Program (YunLing Scholar Project), the Yunnan Province Foundation (Grant No. 202301AU070160), the CAS ``Light of West China" Program and the China Manned Space Project with No. CMS-CSST-2021-A06. Y.C. is supported by the Training Program for Talents in Xingdian, Yunnan Province (Grant No. 2081450001) and the National Natural Science Foundation of China (Grant No. 12203028). J.X.W was supported by the National Natural Science Foundation of China (Grant Nos. 12033006, and 12192221).

Based on observations obtained with the Samuel Oschin Telescope 48-inch and the 60-inch Telescope at the Palomar Observatory as part of the Zwicky Transient Facility project \citep{ZTF-DOI}. ZTF is supported by the National Science Foundation under Grants No. AST-1440341 and AST-2034437 and a collaboration including current partners Caltech, IPAC, the Weizmann Institute for Science, the Oskar Klein Center at Stockholm University, the University of Maryland, Deutsches Elektronen-Synchrotron and Humboldt University, the TANGO Consortium of Taiwan, the University of Wisconsin at Milwaukee, Trinity College Dublin, Lawrence Livermore National Laboratories, IN2P3, University of Warwick, Ruhr University Bochum, Northwestern University and former partners the University of Washington, Los Alamos National Laboratories, and Lawrence Berkeley National Laboratories. Operations are conducted by COO, IPAC, and UW.

\vspace{5mm}
\facilities{ZTF, Fermi LAT}

\software{corner.py \citep{Foreman-Mackey2016}, celerite \citep{Foreman2017}, emcee \citep{Foreman2013}, NumPy \citep{Harris2020}, Matplotlib \citep{Hunter2007}, Astropy \citep{Astropy2022}, SciPy \citep{Virtanen2020}, Pandas \citep{mckinney2010,reback2020}, Statsmodels \citep{seabold2010}, Seaborn \citep{Waskom2021}.
          }

\bibliography{sample631}
\bibliographystyle{aasjournal}


\appendix 
Fig. A1 presents the power spectral density (PSD) of the damped random walk (DRW), corresponding to the result shown in Fig. 2. The fitting results of the DRW model using Gaussian Processes (GPs) are illustrated in Fig. A2. The relationship between the intrinsic variability damping timescale and the Eddington ratio for jetted AGNs is depicted in Fig. A3. Finally, Fig. A4 shows the intrinsic $\gamma$-ray luminosity as a function of the residuals for jetted AGNs.

\setcounter{figure}{0} 
\makeatletter 
\renewcommand{\thefigure}{A\@arabic\c@figure}
\makeatother

\begin{figure}
    \centering
    \includegraphics[width=0.7\textwidth]{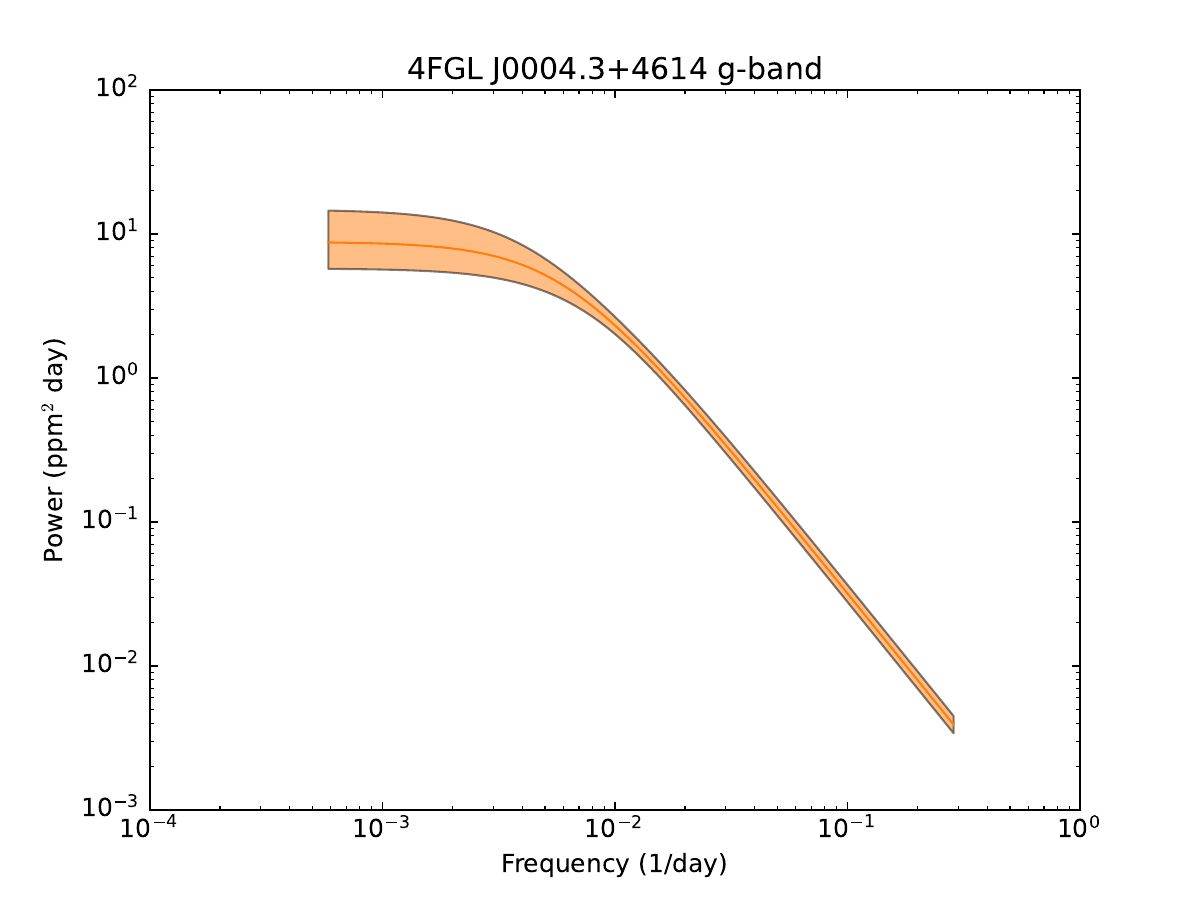}
    \caption{The PSD of DRW corresponding to the result in Fig. 2. The shaded region represents the 1$\sigma$ confidence interval.}
\end{figure}

\begin{figure}
    \centering
    \gridline{\fig{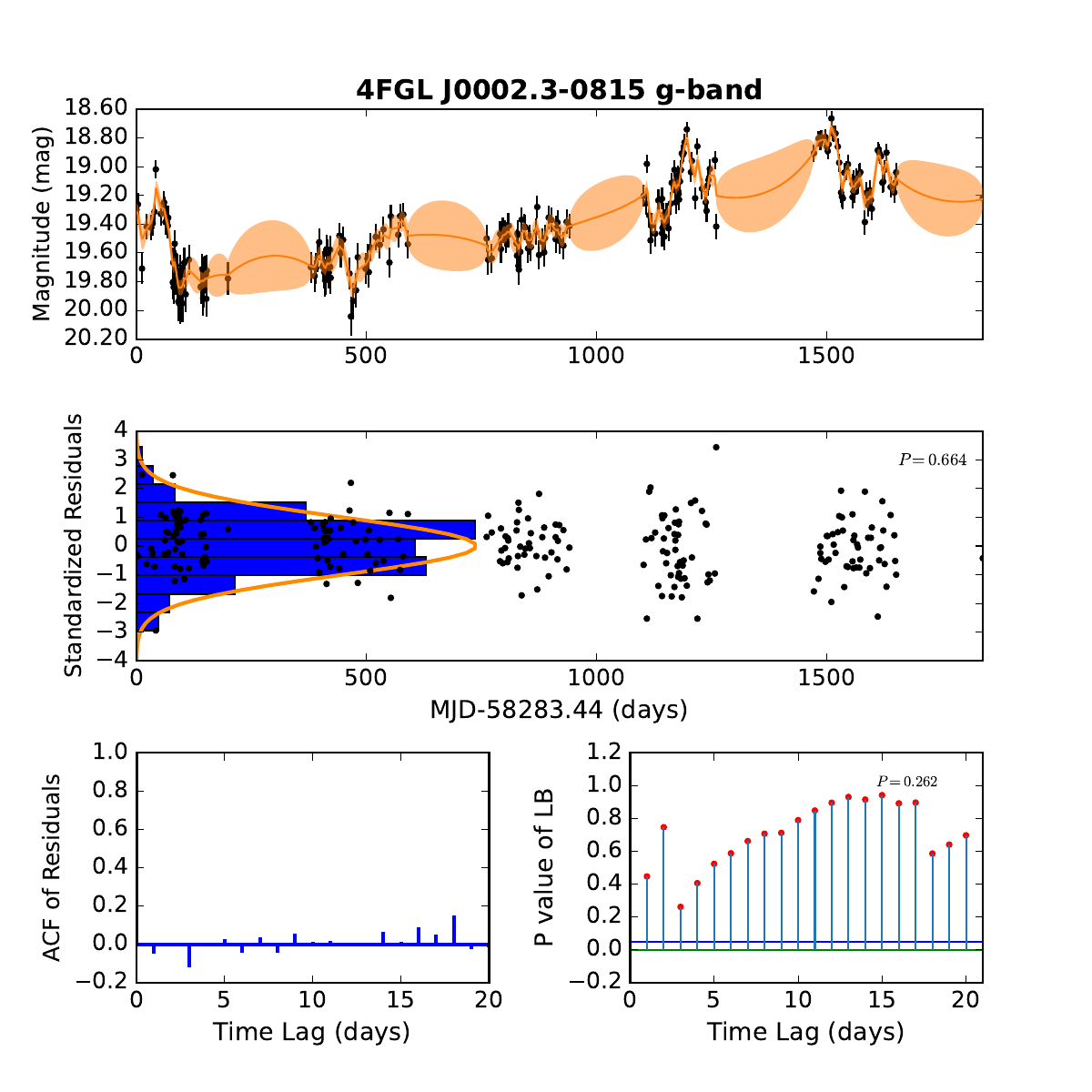}{0.45\textwidth}{(a) An example of a good fit with that the ACF value at a specific lag slightly exceeds the 95\% confidence interval for white noise.}
          \fig{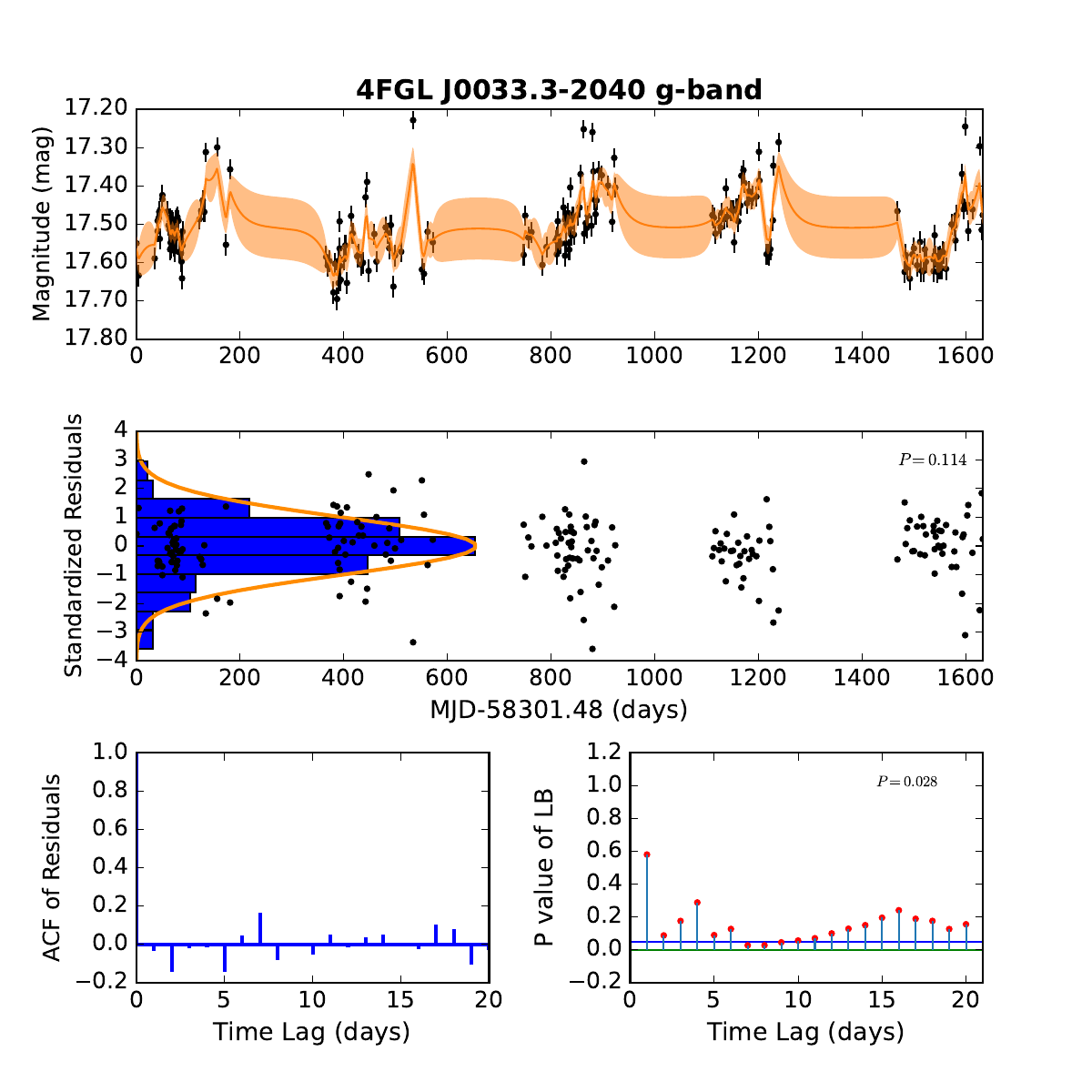}{0.45\textwidth}{(b) An example of a good fit with that the $p$ values from the LB test for a small number of time lags are slightly below 0.05.}}\label{fig:1b}
    \gridline{\fig{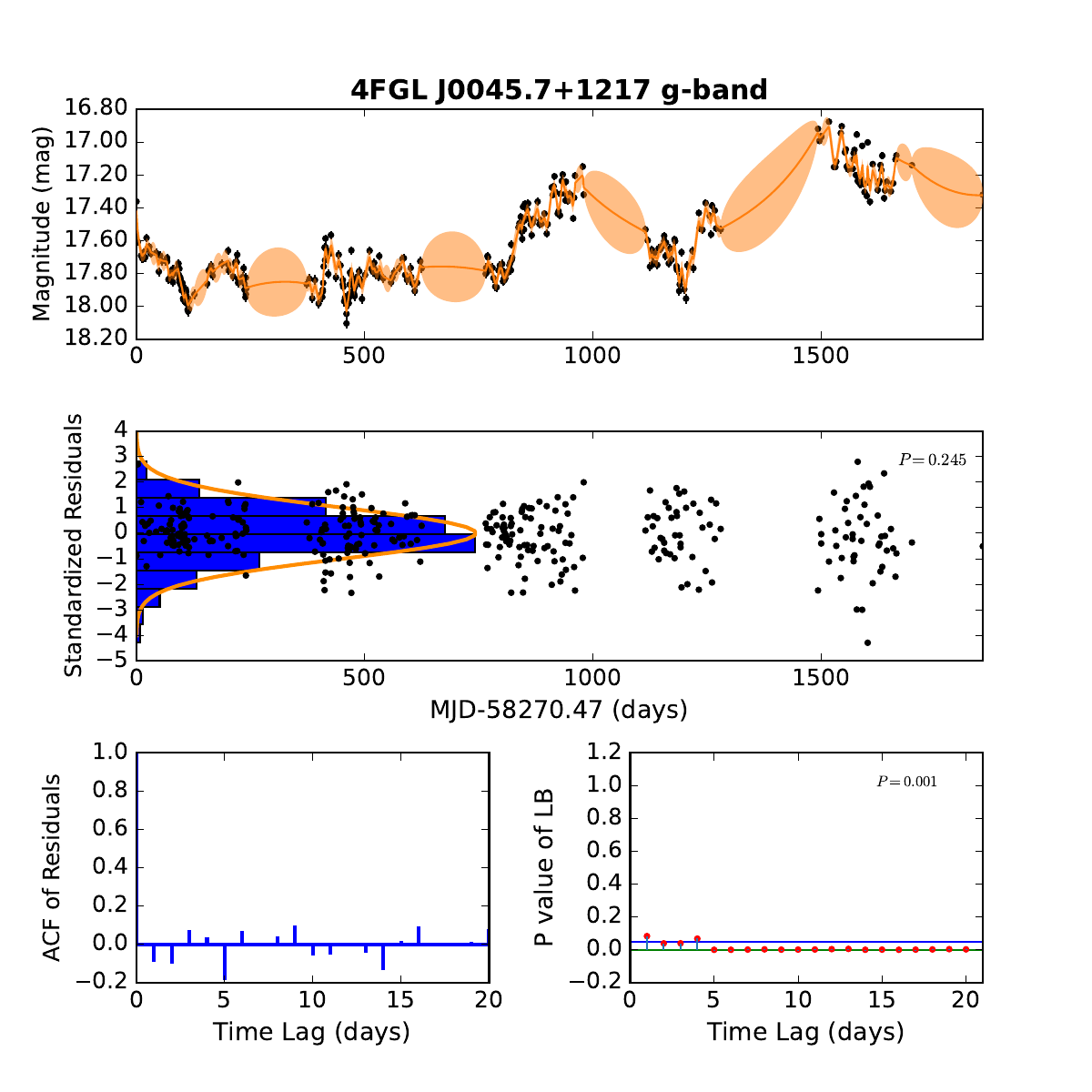}{0.45\textwidth}{(c) An example of a poor fit with that the standardized residuals do not conform to the criteria for white noise.}\label{fig:1c}
          \fig{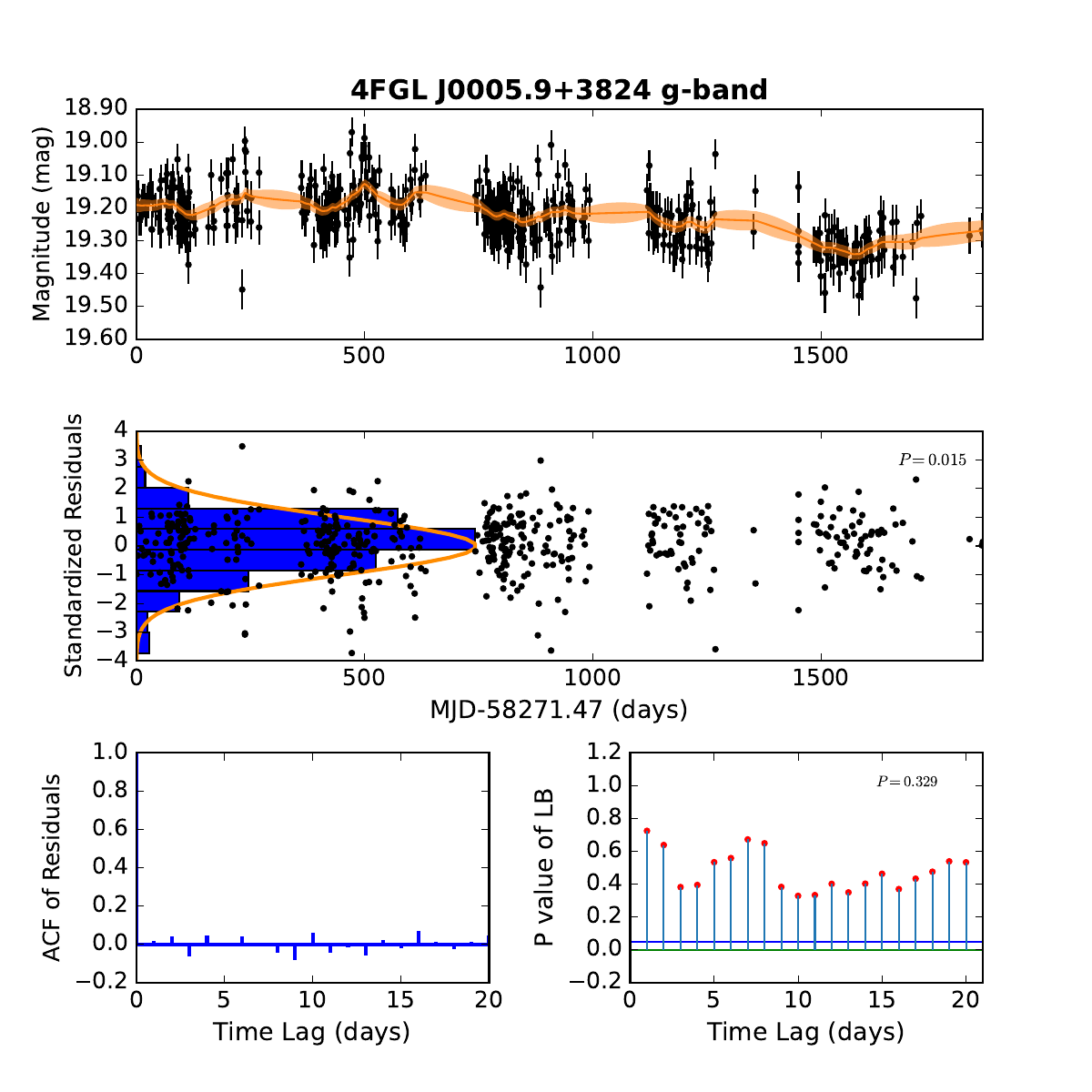}{0.45\textwidth}{(d) An example of a poor fit with that the distribution of the residuals is not normal.}}\label{fig:1d}
     \caption{Examples of model fitting and result. The meanings of the symbols are consistent with those used in Fig. 2a.}
\end{figure} 

\begin{figure*}
\gridline{\fig{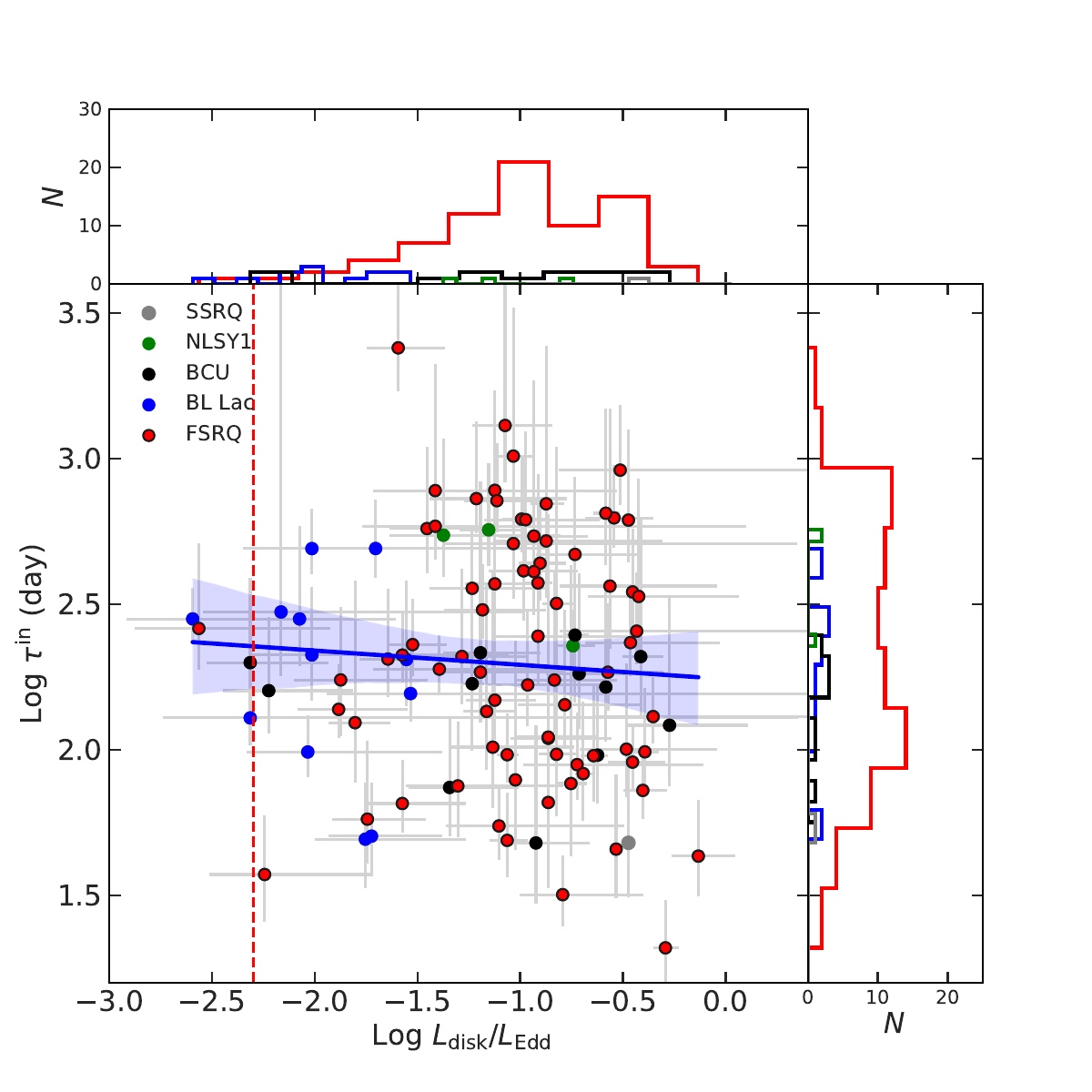}{0.5\textwidth}{(a) The efficient accretion AGNs with good fits of DRW model.}
          \fig{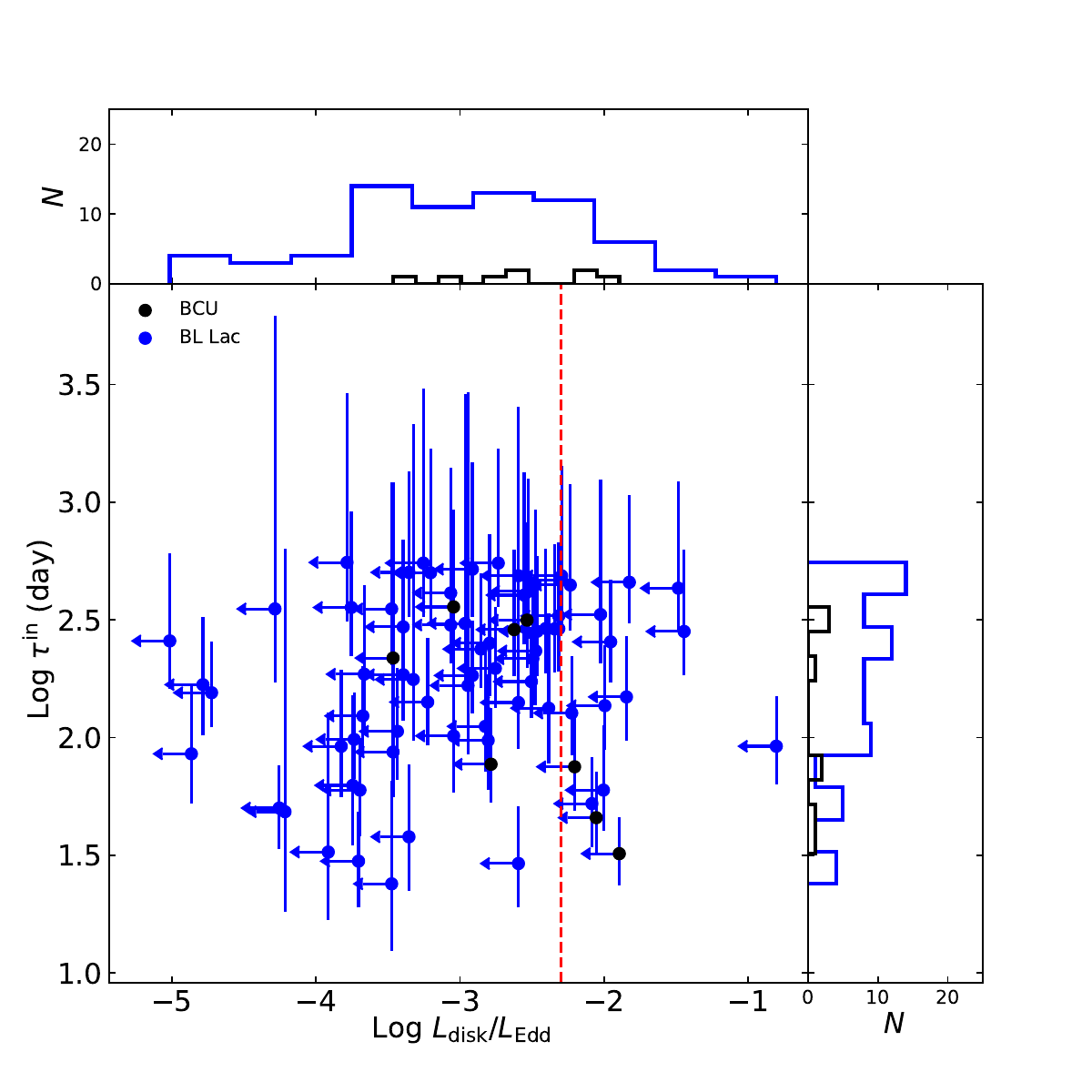}{0.5\textwidth}{(b) The inefficient accretion AGNs with good fits of DRW model.}}
    \caption{The relationship between intrinsic variability damping timescale and the Eddington ratio for jetted AGNs. The meanings of the symbols are consistent with those in Fig. 8, with the exception that the red vertical dashed lines represent the dividing line between efficient and inefficient accretion AGNs. Additionally, the upper limit of the Eddington ratio is indicated in the right panel.}
\end{figure*} 

\begin{figure*}
\gridline{\fig{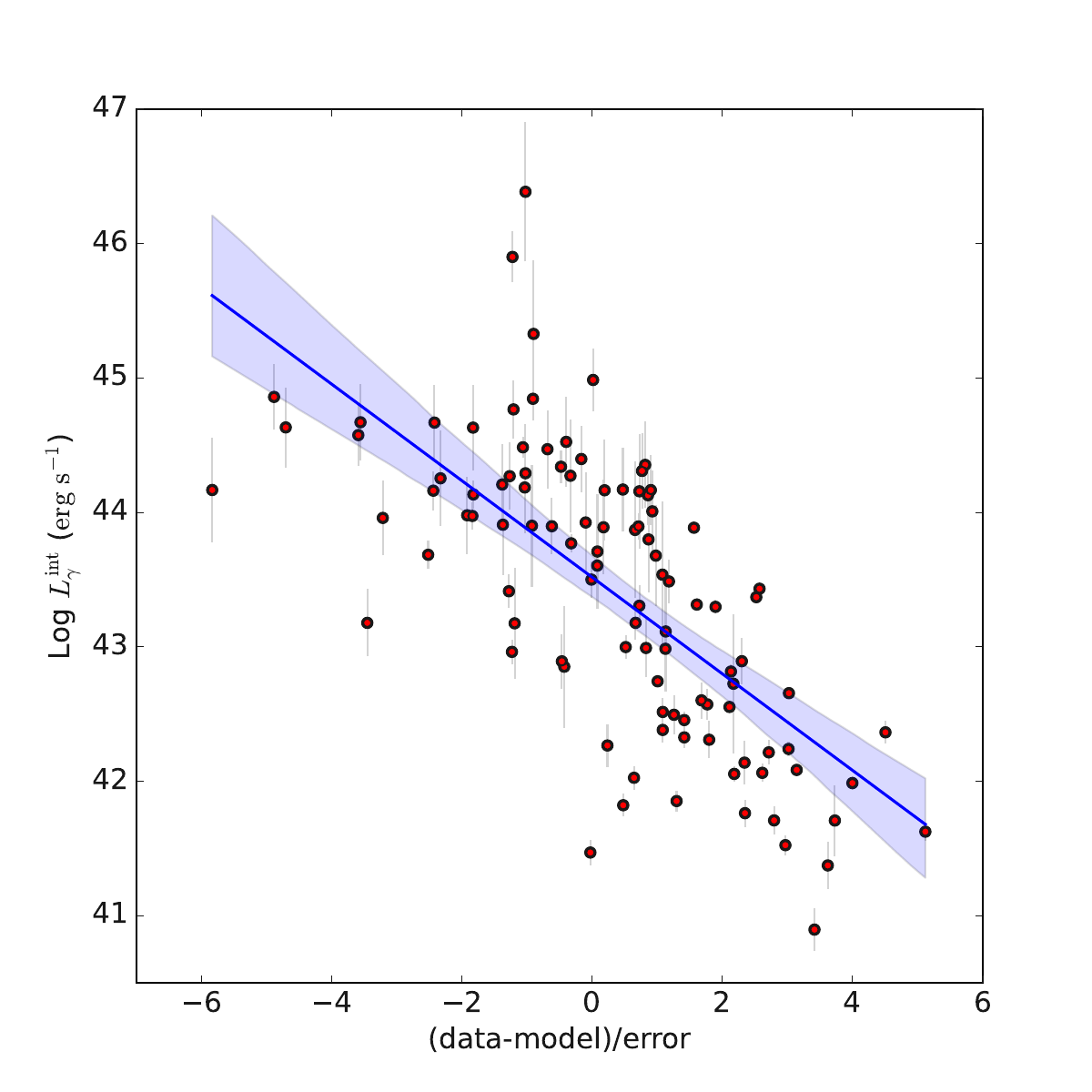}{0.5\textwidth}{(a) The efficient accretion AGNs with good fits of DRW model}
          \fig{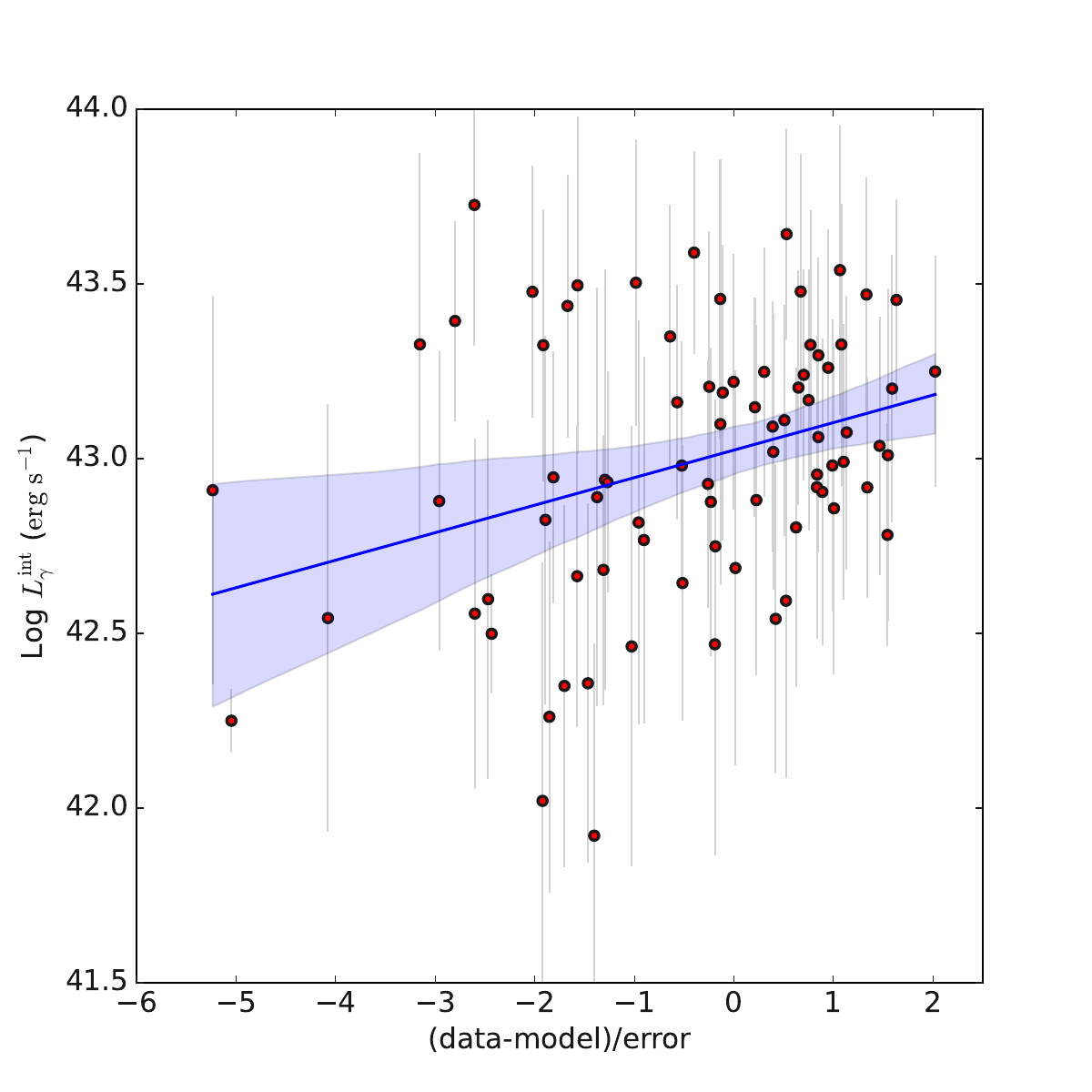}{0.5\textwidth}{(b) The inefficient accretion AGNs with good fits of DRW model}}
    \caption{The relationship between intrinsic $\gamma$-ray luminosity and the residuals for jetted AGNs. These residuals, obtained by subtracting the best-fitting linear model for non-jetted AGNs from the data, correspond to the Fig. 7. The coefficient of correlation is $r = -0.69$ with a significance level of $P = 8 \times 10^{-16}$ for efficient accretion AGNs, whereas for inefficient accretion AGNs, the correlation coefficient is $r = 0.32$ with $P = 0.004$.}
\end{figure*}

\end{document}